\documentclass[a4paper,12pt]{article}

\usepackage{amsmath}
\usepackage{latexsym}
\usepackage{cite}
\usepackage{graphicx}
\usepackage[small,bf,hang]{caption}

\newcommand{\cg}{C(G)}

\newcommand{\first}{\ensuremath{\mbox{1}^{\mbox{\tiny st}}}}

\newcommand{\cc}{^*}

\newcommand{\lomega}{\bra{\Omega}}

\newcommand{\medsp}{\\[0.7ex]}

\newcommand{\nf}{\ensuremath{N_f} }

\newcommand{\romega}{\ket{\Omega}}
\newcommand{\second}{\ensuremath{\mbox{2}^{\mbox{\tiny nd}}}}

\newcommand{\su}{\mbox{\slshape SU}}
\newcommand{\suppot}{\mathcal{W}}

\newcommand{\uone}{\mbox{\slshape U}(1)}

\newcommand{\Ltext}[1]{\ensuremath{\itindex{\mathcal{L}}{#1}}}

\newcommand{\bra}[1]{\langle #1 |}
\newcommand{\dega}{\ensuremath{^\dag}}

\newcommand{\diff}[1]{\mbox{d}#1}

\newcommand{\half}[1]{\ensuremath{\frac{#1}{2}}}
\newcommand{\intd}[1]{\int \!\! #1 \;}
\newcommand{\inv}[1]{\ensuremath{\frac{1}{#1}}}
\newcommand{\ket}[1]{| #1 \rangle}
\newcommand{\metr}[1][]{g_{\varphi \bar{\varphi} #1}}

\newcommand{\pathint}[1]{\int \mathcal{D} #1 \;}

\newcommand{\wphi}[1][]{\suppot_{,\varphi #1}}

\newcommand{\itindex}[2]{\ensuremath{#1_{\mbox{\scriptsize{\itshape #2}}}}}

\newcommand{\varfrac}[2][]{\frac{\delta #1}{\delta #2}}

\newcommand{\nddn}[3][ ]{\ensuremath{#2_{{#1} #3}}}
\newcommand{\ndup}[3][ ]{\ensuremath{#2^{#3}_{#1}}}

\DeclareMathOperator{\re}{Re}
\DeclareMathOperator{\im}{Im}
\DeclareMathOperator{\tr}{Tr}
\DeclareMathOperator{\hc}{h.c.}

\begin{document}
\bibliographystyle{phaip}
%
%
\renewcommand{\thefootnote}{\fnsymbol{footnote}}
\thispagestyle{empty}
\begin{titlepage}

\vspace*{-1cm}
\hfill \parbox{3.5cm}{BUTP-2001/04 \\ hep-th/0102005}
\vspace*{0.5cm}

\begin{center}
  {\Large {\bf \hspace*{-0.2cm}Dynamical Symmetry Breaking
      \protect\vspace*{0.2cm} \\
      \protect\hspace*{0.2cm} in SYM Theories as a 
      \protect\vspace*{0.4cm} \\      
      \protect\hspace*{0.2cm} Non-Semiclassical Effect}\large\footnote{Work
      supported in part by the Schweizerischer Nationalfonds.}}
  \vspace*{2.5cm} \\

{\bf
    L. Bergamin\footnote{email: bergamin@tph.tuwien.ac.at, phone: +43  1 58801 13622, fax: +43 1 58801 13699}} \\
    Institute for Theoretical Physics \\
    Technical University of Vienna \\
    Wiedner Hauptstr.\ 8-10 \\
    A-1040 Vienna, Austria
   \vspace*{0.8cm} \\  

\today

\vspace*{2.3cm}

\begin{abstract}
\noindent
We study supersymmetry breaking effects in $N=1$ SYM from the point of view of
quantum effective actions. Restrictions on the geometry of the effective
potential from superspace are known to be problematic in quantum effective
actions, where explicit supersymmetry breaking can and must be studied. On the other
hand the true ground state can be determined from this effective action,
only. We study whether some parts of superspace geometry are still relevant
for the effective potential and discuss whether the ground states found this
way justify a low energy approximation based on this geometry. The answer to
both questions is negative: Essentially non-semiclassical effects change the
behavior of the auxiliary fields completely and demand for a new
interpretation of superspace geometry. These non-semiclassical effects can break supersymmetry.

\vspace{3mm}
\noindent
{\footnotesize {\it PACS:} 11.30.Pb; 11.15.Kc; 11.15.Tk \newline
{\it keywords:} SYM theory; dynamical supersymmetry breaking; low
energy approximations}
\end{abstract}
\end{center}

\end{titlepage}

\renewcommand{\thefootnote}{\arabic{footnote}}
\setcounter{footnote}{0}
%
\section{Introduction}
The question whether supersymmetry is spontaneously broken or not is of
fundamental importance. Many results concerning this problem have been derived
in the literature. We know that perturbative corrections do not break
supersymmetry. What happens non-perturbatively is not yet clear since there is
no mathematical tool available to describe this regime. Our knowledge about
the behavior of non-Abelian gauge theories is restricted to perturbative
results, semi-classical analysis and simulations on the lattice. But
the ground-state of any non-Abelian gauge theory that is not broken down
completely (up to $\mbox{U}(1)$ factors) is characterized by non-perturbative
effects. Supersymmetry does not help us in this situation, on the contrary:
Many supersymmetric models (e.g.\ $N=2$ SYM) are completely unacceptable in
the perturbative region, as the perturbative $\beta$ function develops a
Landau pole. Exploring the non-perturbative region is not at all simpler than
in ordinary QCD: measurements from the lattice are not yet available as the
Euclidean formulation of the theory is very difficult.

Besides many other models the question of dynamical supersymmetry breaking has been
answered for $N=1$ SYM using the Witten index \cite{witten82} and low energy effective Lagrangian
calculations \cite{veneziano82}. Different Instanton calculations
\cite{amati88,shifman88,kovner97} as well as the concept of Wilsonian low energy effective
actions \cite{shifman86} agree with the scenario of unbroken supersymmetry. But all these calculations have a conceptual problem
in common: Supersymmetry breaking as a hysteresis effect cannot be studied, as
explicit supersymmetry breaking is impossible to include (the notion of hysteresis effects in quantum
field theories is discussed in section \ref{sec:nonpft}). Consequently many
spontaneous effects have to be introduced by assumption, which could
anticipate (non-)existence of dynamical supersymmetry breaking. The assumption
of unbroken supersymmetry has important influence
on the picture of superstring theory: Directly as in certain limits flat
field theories appear on the branes of string theories and indirectly by
establishing duality arguments within supersymmetry.

In this work we want to re-discuss the question of dynamical supersymmetry
breaking from a very fundamental point of view. First we introduce the basic
concepts used to determine the vacuum structure of quantum field theories
(hysteresis effects) and
apply them to $N=1$ SYM. Such hysteresis effects have already been studied for $N=1$ and $N=2$ SYM in
\cite{bib:mamink,bergamin00:1}. But these results are incomplete as well: First it
is difficult to compare them with other calculations as a non-standard
structure of the QCD vacuum plays an important role and second the geometric
approach used therein is known to be problematic in quantum effective actions.

The second part of the paper is devoted to the discussion of this last point:
the relevance of superspace geometry in the context of quantum effective
actions and its connection to other formulations as Wilsonian low energy
effective actions or effective Lagrangians. It is of main importance
as from our point of view only the quantum effective action can tell us the correct ground state. A
justification of the geometric approach for effective Lagrangians or Wilsonian
effective actions must thus be derived therefrom. Our
result suggests that this cannot be done: A consistent description of the
quantum effective action is found together with non-semiclassical effects, only.
These non-semiclassical effects can break supersymmetry and demand for a new
interpretation of superspace geometry. An alternative scenario is possible: Supersymmetry is unbroken but
has a phase transition in the variation of the gluino mass at $m=0$. We
suspect that the resulting theory is highly infrared sick and probably does not exist at all.
Of course these results are not free of assumptions either. As exact results
are not obtainable we follow the philosophy of semi-classical analysis as far
as needed to be able to make any statements. Especially we assume that we can
define composite operators and their vacuum expectation values and that the
unphysical fields from quantization need not be included explicitly.

In this paper we mainly want to collect known results from non-perturbative
supersymmetry as well as non-perturbative QCD and classify them from the point
of view of the fundamental question: ``How do we find the true ground-state of
a quantum field theory?'' We propose a new scenario for the dynamics of SYM,
but this result is rather speculative. First steps towards a concrete model of
this type will be presented elsewhere \cite{bergamin02:1}.

The paper is organized as follows: In section two we review some basic facts
about non-perturbative field theories and hysteresis effects. Here we
formulate our recipe how to find the true ground-state of a quantum field
theory. Section three applies these ideas to $N=1$ SYM and compares our ansatz
with other low energy approximations known in literature. In section four constraints on
dynamical supersymmetry breaking independent of our approach are discussed. It
is shown in section five that the standard interpretation of SYM theory is not
compatible with all constraints on the dynamics found so far. A (rather
speculative) way out of this is developed (section five) and discussed in section six. Therein
we also comment on the Witten index and on alternative approaches to the low
energy dynamics of SYM. Finally we give some comments about more
complicated models (section seven) and draw our conclusions (section eight).


\section{Non-perturbative QFT as Thermodynamical Limit}
\label{sec:nonpft}
In this section we want to review some basic aspects of non-perturbative
quantum field theory and discuss its relevance for a modern approach to a 4-d
QFT, where exact calculations in the non-perturbative sector are
not available. Physical amplitudes are derived from the generating functional of
the Greens functions $Z_M$ or from the heat kernel $Z_E$, written as
path integrals:
  \begin{align}
    Z_M[J] &= \pathint{\phi} e^{i S_M(\phi,J)} & Z_E[J] &= \pathint{\phi} e^{-
    S_E(\phi,J)}
  \end{align}
where $S_E$ is the Wick rotated action of $S_M$. At least formally we can define to every source extension of a classical action an effective action and
an effective potential. These quantities are obtained by a Legendre
transformation with respect to the sources and the new variables are called
classical fields:
\begin{subequations}
\label{eq:effact}
\begin{align}
  Z_M[J] &= e^{i W[J]} & \varfrac[W{[}J{]}]{J(x)} &= \tilde{J} = \lomega \phi(x)
  \romega_J \medsp
  \Gamma[\tilde{J}] &= \intd{\diff{^4x}} \bigl( J(x) \varfrac[W{[}J{]}]{J(x)}
  \bigr) - W[J] & \itindex{V}{eff} &= \inv{V^4} \bigl. \Gamma[\tilde{J}] \bigr|_{p
  \rightarrow 0}
\end{align}
\end{subequations}
The above definitions are given in Minkowski space, analogue definitions for
Euclidean space follow straightforwardly. In the last equation $V^4$ is the
space-time volume and $\phi$
are the operators associated with the sources $J$. They may be basic
field operators as well as composite ones. In the first case the effective
action is often called 1PI effective action, as it is the generating
functional of the 1PI irreducible graphs. Not every source extension can lead
to a well defined effective action. The mapping $J \rightarrow \tilde{J}$ must
be one-to-one and consequently the effective potential is always a convex
function in the classical fields. We will discuss some of these
problems when defining our supersymmetric effective action. If $\Gamma[J]$
exists, it defines thermodynamical relations between the sources
and the spontaneous parameters of the associated operators. In the limit $V^4
\rightarrow \infty$ we have to turn off all sources in order to re-obtain our
original theory and thus
\begin{align}
  \varfrac{\tilde{J}(x)} \Gamma[\tilde{J}] &= J(x) \rightarrow 0 &
  \varfrac{J(x)} W[J] &= \tilde{J} \rightarrow \tilde{J}\cc
\end{align}
A non-vanishing value of $\tilde{J}\cc$ indicates the appearance of a
spontaneous parameter (vacuum expectation value).

While the meaning of the path
integral is well understood in perturbation theory its interpretation in the
non-perturbative sector is not straightforward. As we are trying to derive
the properties of a quantum field theory from a classical expression like the
action we can use the results of constructive field theory as a guide (see
e.g. \cite{glimm85:1,glimm85:2} for a review of constructive quantum field
theory). It is well known that we can solve a theory by introducing an UV and
an IR regularization, only. Many examples of perturbative and non-perturbative
UV regularization schemes are known and it is an important feature of a well
defined theory that the resulting dynamics are independent of the
regularization scheme. The issue of renormalization in the UV region is not
topic of our work. Rather we assume that it can be performed not only in
perturbation theory but also in the non-perturbative region. Especially we
assume that the definition of an operator in perturbation theory can be
extended to the non-perturbative region and that we can give an interpretation
of its vacuum expectation value.

The IR regularization plays a quite different role: In
the context of perturbation theory a complete interpretation has been given by
Bloch and Nordsieck \cite{bloch37} and by Weinberg \cite{weinberg65}, while it
plays an essentially different role in non-perturbative dynamics, which we want to discuss in the
remainder of this section. How should we choose the IR regularization? If the
classical theory has a mass gap we only need to restrict the trilinear and cubic
interactions to a compact support, else (as in gauge theories) we essentially have to put the
theory into a finite volume. In the latter case we are confronted with the problem
of choosing boundary conditions (BC's). We further have to distinguish fermions
from bosons:
\begin{itemize}
\item The fermionic path integral is defined as the functional determinant of
  the corresponding operator. Therefore the eigenvalue problem has to be
  studied, which makes it understandable that even the infinite volume limit
  can depend on the BC's. An additional problem appears when regularizing the theory on topological
  spaces. Zero modes (Instantons) then appear and the functional determinant
  is zero if at least one fermion is massless. We should remember that the
  definition of the partition function is only meaningful when attaching
  sources. Introducing sources $\eta$ and $\bar{\eta}$ for the fundamental
  fermion fields the partition function in presence of $k$ zero modes $\xi_i$
  can be written as
  \begin{equation}
  \label{eq:fermiondet}
    Z[\eta, \bar{\eta}] = \prod_1^k \bra{\bar{\eta}} \xi_i \rangle
    \bra{\bar{\xi}_i} \eta \rangle \exp\bigl\{- \intd{\diff{^4x}\diff{^4y}}
    \bar{\eta}(x) G_e (x,y) \eta(y)\bigr\} {\det}'(i \mathcal{O})
  \end{equation}
  where $G_e(x,y)$ is the Green function on the space orthogonal to the zero
  modes and the prime indicates that the determinant is calculated over the
  non-zero eigenvalues, only. A detailed but rather technical discussion of
  this problem has been given by Rothe and Schroer \cite{rothe80}, see also
  \cite{sachs92}. Simple counting of zero modes shows that the chiral
  condensate of a theory regularized on the sphere or on the torus vanishes
  identically for $\nf > 1$ \cite{leutwyler92,wipf95}. This boundary effect
  does not have any physical meaning for the theory at infinite volume but is
  a wrong choice of IR regularization. The example shows how certain (not even
  exotic but very popular) BC's can lead to wrong conclusions if the
  thermodynamical limit is interpreted too naively. For further discussions
  and possible BC's solving the above problem
  we refer to different studies of the Schwinger model
  \cite{jayewardena88,sachs92,wipf95,durr96}.
\item No simple interpretation of the path integral is available for bosons
  but it is said to be the ''sum over all possible paths''. This is misleading in any model
  with degenerate vacua (e.g.\ due to symmetry breaking). This can be seen in simple
  examples as $\phi^4$ in the Higgs phase or a one-dimensional ferromagnetic
  spin chain. The rotational symmetry is not anomalous and we can choose an UV
  regularization respecting the latter. Choosing BC's respecting the symmetry
  as well we get vanishing expectation values of the scalar field or of the
  spontaneous magnetization
  \begin{equation}
    \lomega \phi \romega = \pathint{\phi} e^{iS} \phi = 0
  \end{equation}
  as the regularized path integral and the action are even under the symmetry
  while the field is odd. The simplest realization of this situation is the
  spin chain regularized on a circle. A straightforward interpretation of this effect
  exists: The theories do not have a single vacuum, but infinitely many
  connected by symmetry transformations. To get a meaningful result one has to avoid integrating
  over all vacua but has to pick out one of them which is done by an
  appropriate choice of ''boundary
  conditions'' \cite{glimm85:2} (we put this in quotation marks to indicate that there are many
  ways to impose such a constraint).
\end{itemize}
Physical interpretations of this behavior can be given: In presence of
spontaneous parameters there exist phase transitions and indeed in this case
boundary terms can get larger than volume terms. Considering the vacuum state
in the limit $V^4 \rightarrow \infty$ the latter must be a pure state while
ground states in the finite volume are in general not pure. Thus the above ''vacua'' of the $\phi^4$
theory are not acceptable. Although formally correct these
interpretations of the phenomena fail to be applicable here: They assume that
we know the vacuum expectation values (vev's) of the Hamiltonian as well as
of the basic field operators (in other words the redefinitions $\mathcal{H}
\rightarrow \mathcal{H}'$ and $\phi \rightarrow \phi'$ such that $\lomega
\mathcal{H}' \romega = 0$ and $\lomega
\phi' \romega = 0$). If this is known we can
choose one such $\phi' = \phi - \lomega \phi
\romega$ and the scalar path integral is extended over the dynamical part
$\phi'$, only. In this work however we would like to use the thermodynamical
limits to \emph{determine} the above shifts and therefore these
interpretations can be given a posteriori, only. Different thermodynamical limits (i.e.\ infinite volume limits starting from
different finite volume preparations of the system) are therefore treated independently although the
final result may give a definite interpretation of the ''wrong'' limits in
terms of the limits actually leading to the correct ground-state. How can we read off the correct
vacuum expectation values? The idea is rather simple: At least one limit (or one
class of limits defined up to symmetry transformations) does lead to the correct ground
state. The latter is defined to be the absolute minimum of the effective
potential. Therefore we have to calculate the limits from all possible
perturbations (we call these states trial vacua $\ket{0}$) and pick out the one(s) that
minimize the effective potential:
\begin{align}
  \romega &= \min_{\itindex{V}{eff}} \{ \ket{0}\}
\end{align}
We should give some additional comments: We noted above that the effective
potential is a convex function and thus there can exist only one minimum
thereof. This is true for one particular set of sources and BC's. Here we are
speaking about \emph{all possible} perturbations and they can lead in principle to
infinitely many different convex functions, each having some minimum. These
different minima need not be physically equivalent but are exactly the trial
vacua $\ket{0}$. In this generalized sense the effective potential (being
the set of all possible convex effective potentials) can now have more than
one minimum and the correct ground state is indeed the absolute minimum with
respect to the energy.

The perturbations can be boundary conditions or global sources
\cite{minkowski90,bib:mamink,bib:markus,bergamin00:1}. In the latter case sources are no
longer seen as spatial restricted perturbations (typically as $\delta$
functions) but are extended over a large part of space-time, though the
boundary condition $J(x) \rightarrow 0$ for $x \rightarrow \infty$ still
holds. The sources can then be seen as new coupling or mass parameters of
the theory that have the following special features: First their value is not fixed (to obtain the original theory they have to
be turned off in the end) and second an associated classical field can be defined. A
simple realization can be seen as follows \cite{bib:mamink,bib:markus}: We
split the finite volume $V$ into a part $\itindex{V}{sub}$ and $V \setminus
\itindex{V}{sub}$.  $V \setminus
\itindex{V}{sub}$ shall contain the whole boundary. Now we can choose the
sources non-vanishing but constant inside $\itindex{V}{sub}$ and vanishing
outside. The thermodynamical limit is taken as $\itindex{V}{sub} \subset V
\rightarrow \infty$ and the sources are getting relaxed in the end.

Using the concept of global sources the effective action can depend formally
on both, the sources and the classical fields: In the above limit the sources
are constant $J(x) = J_0$ in (almost) the whole space-time and the procedure can be seen as
a choice of new boundary conditions $J(x) \rightarrow J_0$ ($x\rightarrow
\infty$). The variation with respect to the source is then a small
perturbation on its constant part. Defining $J = J_0 + \Delta
J(x)$, $J_0$ is irrelevant in the process of the variation
$\varfrac[W{[J]}]{J(x)} = \tilde{J}(x)$. But of course the classical field
itself depends on the value of $J_0$. We can then see the effective action as
a function of the classical field and the constant source $\Gamma =
\Gamma[\tilde{J};J_0]$ where $\tilde{J}$ itself depends on $J_0$. Of course
this is much the same as introducing a coupling constant $J_0$ and attaching a
local source $\Delta J$ to the same operator as $J_0$ and the dependence on the static
part of the source plays a similar role as the dependence of the effective
action on any coupling constant. Nevertheless we use the notion of global
sources to keep track of the role of all parameters.

Global sources are especially useful if they generate a classical mass-gap inside the perturbed region and
make BC's (at least for the corresponding fields) irrelevant. As an example
the above mentioned problem of vanishing chiral condensates can be solved
without introducing different BC's but by attaching a global source having the
effect of a fermion mass.

We should be careful with the limits of this prescription: We can always find the correct minima
of the potential but the corresponding effective action needs not describe the
correct dynamics. This can e.g.\ happen if a theory with instanton-like effects
is perturbed in such a way that all or some instantons are getting suppressed. Our theory may be non-Hamiltonian, too. In
this case we will find more than one ground state (up to symmetries) and the
actually chosen state will depend on an external parameter.

Of course this program can be realized in principle, only. But there is a
simple way to extract the relevant perturbations: To make the breaking of a
symmetry visible, we need a trigger of the latter in the IR regularized
theory. Therefore the interesting perturbations break the symmetries in
question. In analogy to a spin-system we call a spontaneous parameter,
associated with such a perturbation, hysteresis effect. Further restrictions on
the choice of sources arise from the renormalization procedure:
Symmetry invariance or covariance of the classical system including all sources must be
extendible to the quantum theory. Else the corresponding symmetries are not
realized on the level of the effective action. Moreover the perturbation must
hold two stability conditions. First the resulting trial vacuum should be stable in a renormalization group
analysis. This means that the classically trivial relaxing of the sources has
a meaning in quantum theory: when turning the classical sources off the
quantum system tends towards (and finally reaches) the original system. In
addition a
Minkowskian theory can have unstable potentials. Indeed the
vacuum-to-vacuum transition probability $|\bra{\Omega(t = + \infty)} \Omega(t = - \infty) \rangle|^2 =
|e^{iW}|^2 = e^{- 2 \im W}$ is $< 1$ if the effective
potential is complex leading to a decay of the vacuum. Thus the effective
potential must be real at any point of the perturbed system. If (at least in
some range of the source parameters) the above
conditions are met and the renormalized quantities stand in a one-to-one
correspondence to the classical ones we can freely replace (in that range) the classical
parameters by the renormalized ones.

As an application of this principle we note that it determines the value of
$\theta$ in any QCD-like theory uniquely. This has been studied in detail in
the Appendix of \cite{bergamin00:1} using the Instanton picture.  We will see that
$S^4$ or $T^4$ as regularization spaces of the heat kernel or the generating
functional of SYM are not sensitive to supersymmetry breaking. However
the arguments given in \cite{bergamin00:1} straightforwardly extend to any
regularization space where $\theta$ has a non-trivial meaning as well as to
perturbations by fermion masses (note that you have to introduce sources to both operators
$\bar{\psi} \psi$ and $i \bar{\psi} \gamma_5 \psi$). This is especially true
when using local boundary conditions (e.g.\ ''bag'' BC's \cite{wipf95,durr96}) or
in a quantization on the light cone \cite{cartor00}.


\section{The Effective Action of $\mathbf{N=1}$ SYM}
We want to apply the program sketched above to $N=1$ SYM. The
Lagrangian is given by
\begin{align}
\label{eq:classlag}
  \mathcal{L} &= \inv{8 \cg} \bigl(\intd{\diff{^2 \theta}} \tau \tr W^\alpha
  W_\alpha + \hc \bigr) + \Ltext{GF} + \Ltext{ghost} & W_\alpha &= - \bar{D}^2
  (e^{-V} D_\alpha e^V)
\end{align}
with the prepotential $V$ used to quantize the theory in superspace. We work
in Minkowski-space with the generating functional
\begin{align}
  Z[J] = \pathint{\phi} e^{i \bigl(S_0(\phi) + S_J(\phi,J)\bigr)}
\end{align}
To decide whether supersymmetry is broken dynamically or not we introduce a
set of global sources that
\begin{itemize}
\item break supersymmetry as well as chiral symmetry,
\item connect the supersymmetric theory with some configuration where other
  dynamical effects (confinement, glue-ball) are (though not understood) well
  accepted,
\item could still be sensitive to the special geometry of supersymmetric theories.
\end{itemize}
The above conditions are satisfied by the concept of local couplings, where
the coupling constant is replaced by a chiral superfield \cite{bib:mamink}. We define a quantum effective action
\begin{align}
  \Gamma[\tilde{J},\Tilde{\Bar{J}}] &= \intd{\diff{^4x}} \bigl( J(x) \varfrac[W{[}J{]}]{J(x)} +
  \hc \bigr) - W[J,\bar{J}]
\end{align}
where $J = \tau + \theta \eta - 2 \theta^2 m$ is the local coupling
superfield. $\Gamma[\tilde{J},\Tilde{\Bar{J}}]$ and $W[J, \bar{J}]$ are connected by
thermodynamical equilibrium conditions and in the thermodynamical limit the
effective action obeys the (anomalous) Ward-Identities
\cite{bib:mamink,bergamin00:1}. The chiral source field defines a set of
three dual
fields $\tilde{J}$. Its components are the vev of the Lagrangian, of the gluon
condensate and of a spinor, which represents the goldstino in case of broken supersymmetry. The following assumptions have to be made to be able to discuss
supersymmetry breaking in a similar way to Veneziano and Yankielowicz
\cite{veneziano82}: The above effective action exists at least in its static
limit and therein the classical fields $\tilde{J}$ can be re-combined to a chiral
superfield obeying the standard supersymmetry transformation rules. We would
like to make some comments on this:
\begin{itemize}
\item The gluino condensate is certainly a natural perturbation to study
  dynamical supersymmetry breaking . The latter is expected to be connected
  to other dynamical effects of which chiral symmetry breaking is the only one
  accessible directly. Nevertheless other or additional breaking terms can be
  introduced at the classical level. Renormalization group analysis  however suggests that
  such hard supersymmetry breaking terms are forbidden due to instabilities of the
  supersymmetric solution
  \cite{oehme84,oehme85:1,oehme85:2,zimmermann85,oehme86}. Although this is
  not of main interest in this context we would like to note that the same is
  true for possible gauge symmetry breaking terms
  \cite{kraus91:1,kraus91:2,kraus91:3}.
\item Once we have identified the gluino term as the only reasonable perturbation,
  we can try to construct a chiral field from the classical variables having
  all the properties required. In
  perturbation theory such a field, the anomaly multiplet, in fact exists. It
  has been constructed in the Wess-Zumino model and in SQED and its existence has been proven in the
  non-Abelian case \cite{clark78,clark80,piguet82:1,piguet82:2}.
\item In perturbation theory $N=1$ SYM with local coupling constant has been studied
  recently \cite{kraus01,kraus01:3}. The author finds an anomalous breaking of
  supersymmetry, as the conditions
  \begin{align}
    \mathcal{S}(\Gamma) &= 0 & \intd{\diff{^4 x}} (\varfrac{\tau} -
    \varfrac{\bar{\tau}}) \Gamma &= 0
  \end{align}
  cannot be satisfied simultaneously if the coupling is space-time
  dependent ($\mathcal{S}$ denotes the Slavnov-Taylor operator), but there
  appears an anomaly in one of the above identities. If the anomaly is put
  into the Slavnov-Taylor operator the simple notion of superfields is
  lost. But we can put the anomaly into the $(\tau - \bar{\tau})$-identity as
  well. Then superspace is still valid and we can expect that the effective
  action is an integral over the standard superspace. A more detailed
  discussion of the relevance of this work to the conclusions of this paper
  must be postponed to a future publication.
\end{itemize}
\subsection{Quantum or Wilsonian Effective Action?}
Besides other models $N=1$ and $N=2$ Yang-Mills theories without
\cite{veneziano82,seiberg94} and with matter fields
\cite{taylor83,davis83,affleck84,affleck85,seiberg94:2} have  been studied
using the concept of Wilsonian low-energy
  effective actions (LEEA's) or in the first case of a low-energy effective
  Lagrangian. In combination with instanton calculations these concepts have been extremely successful to explore the non-perturbative
region of supersymmetric gauge-theories. As our comments on and our criticism
of these concepts do not rely on the details of the results we do not want to
repeat them at this place. Besides the original works cited above the results
have been summarized in several review articles and lecture notes, e.g.\ \cite{intriligator96,
  peskin97, shifman97, shifman99, alvarez-gaume97, divecchia98, bilal95}. The motivation to use LEEA's instead of quantum
  effective actions (QEA's) is twofold: The authors would like to have an
  expression local in the fields, representing all relevant dynamics at low
  energies and they
  assume that the superspace can be reconstructed on the level of these fields
  completely.  The
low energy dynamics can then be written as
\begin{equation}
\label{eq:geomefflag}
  \Ltext{eff} = \intd{\diff{^4 \theta}} K(\Phi, \bar{\Phi}, J, \bar{J}, \Lambda) +
  \bigl( \intd{\diff{^2 \theta}} W(\Phi,J,\Lambda) + \hc \bigr)
\end{equation}
where $\Phi$ represents the quantum fields, $J$ the local coupling and
$\Lambda$ the scale of the Wilsonian action. Local couplings and scale explicitly appear in the LEEA
only and are treated as background fields. In both formulations any additional parametrical dependence (not
  expressible as an integral over superspace) is excluded. The main restriction on the above
form is the holomorphic dependence of the superpotential on its fields. Thus
every field appearing in the superpotential must depend on other fields in a
holomorphic way as well. The kinetic term of the gauge fields is usually written as chiral integral
but in contrast to the superpotential the latter is not irreducible and thus
the non-renormalization theorem does not hold for this term. Nevertheless the
effective kinetic part in
the low energy approximation is written as a chiral
integral and the holomorphy restriction is imposed. It follows that the $\beta$-function of
SYM must be a holomorphic function. This condition does not apply to QEA's which should not
surprise, as the renormalization scheme itself does not have such a holomorphy
constraint for the renormalized coupling constant. To escape this problem Shifman and Vainshtein introduced the notion of
Wilsonian low energy effective actions within supersymmetry
\cite{shifman86,shifman91}. Indeed the coupling constant of the LEEA differs
from the one of the QEA by renormalization effects. The authors come to the
conclusion that these effects turn the non-holomorphic coupling constant of
the QEA into a holomorphic one of the LEEA. Further discussions of this effect
have been given by Dine and Shirman \cite{dine94}. The same
result was obtained by Arkani-Hamed and Murayama \cite{arkani-hamed97} using a
different picture than Shifman and Vainshtein. Although we do not agree with
the treatment of the vacuum angle that serves as an example for $N=1$ SYM in
\cite{shifman91} we insist that non-locality and the non-holomorphic dependence
are crucial characteristics of QEA's.

The first important observation leading to results different from the ones
cited above is the following: In our opinion we have to adjust the
construction principle to the QEA and not the other way around, though a semi-classical ansatz for the
QEA may be more difficult to find. As pointed out in section \ref{sec:nonpft}
we have to study the hysteresis curve of explicitly broken supersymmetry back to
the supersymmetric point and the natural formulation of this program is the QEA
while the above described LEEA does not help us in this situation (though the
low energy effective Lagrangian of Veneziano and Yankielowicz is conceptually
different from the Wilsonian LEEA it suffers from the same problem; this will
become clear in the discussion of section \ref{sec:auxbreaking}). Whether
there exists a holomorphic coupling constant allowing the formulation in form
of a LEEA must be answered after the true ground-state has been found using
the QEA. We should not expect that this is possible: From the perturbative
analysis of SYM with local coupling constant \cite{kraus01,kraus01:3} it has been found that the origin
of the non-holomorphic dependence of the $\beta$-function is essentially
different from the propositions in \cite{shifman86,shifman91} and \cite{arkani-hamed97} and that a simple
redefinition of the coupling constant as proposed in the latter works cannot
lead to a holomorphic coupling constant. In the language of
\cite{kraus01,kraus01:3} Shifman and Vainshtein assume that the LEEA can be
formulated using invariant counterterms, only. Indeed in this case the
$\beta$-function is strictly 1-loop and the coupling constant holomorphic. But
it is not evident, why the invariant
counterterms should play a preferred role. Shifman and Vainshtein argue that all effects from
non-invariant counterterms are IR-divergences and are thus regularized in the
LEEA. E.\ Kraus does not come to the same conclusion.

Let us assume for a
moment that the point of view by Shifman and Vainshtein is correct. Why should
we then look at the LEEA instead of the QEA?  Following Shifman and Vainshtein we should look at the LEEA as this quantity alone is free of infrared
subtleties. In the original paper introducing
this concept \cite{shifman86} this is rather seen as a trick to obtain a
holomorphic $\beta$-function, the LEEA is not seen as a physical object. In
\cite{shifman91} the authors revised this opinion and they concluded that the
objects (esp.\ the value of the coupling constant) from the LEEA are
physical, in contrast to their counterparts from the QEA. From our point of view this is a misunderstanding of
the infrared-problem of these theories. The serious infrared problem in
perturbation theory is (hopefully) an effect of a wrong expansion and is
getting removed in the non-perturbative region by the dynamical formation of a
mass gap, a fact that Shifman and Vainshtein implicitly have to assume as
well. There exist exactly two possibilities for the non-perturbative behavior of
the theory:
\begin{itemize}
\item The IR problem is getting solved. Then we can freely remove any IR
  regulators and there exists no conceptual reason to prefer the LEEA (or any
  other IR-regularized formulation) compared to our QEA, but our discussion
  shows that we are forced to use the QEA (or an equivalent formulation
  including the full dynamics): Such a formulation alone can show how the
  IR-divergences are getting removed, i.e.\ which symmetries survive this
  procedure and which are broken dynamically. By introducing an arbitrary IR
  regulator, Shifman and Vainshtein remove the relevant part of the dynamics
  by hand and thus miss an interesting point in the discussion of
  supersymmetry breaking.
\item The infrared problem is not getting solved. Then indeed the QEA is
  ill-defined but the LEEA is useless as well as the underlying QFT does not
  exist at all. Clearly we have to exclude this possibility by assumption.
\end{itemize}
We can now give a more detailed formulation of the assumptions made for
our QEA:
\begin{itemize}
\item As Veneziano and Yankielowicz we assume that all relevant low energy
  degrees of freedom are represented by the dual fields to $J$, i.e.~the
  Lagrangian itself, the gluino condensate and the would be goldstino in the case
  of spontaneous supersymmetry breaking.
\item We assume that the explicitly broken theory with massive gluinos has a
  low energy behavior similar to QCD and that it does not undergo any phase
  transitions when varying $m$.
\item The effective action defined this way is a supersymmetric extension of
  the 2PI effective action constructed in perturbation theory by Cornwall,
  Jackiw and Tomboulis\footnote{The author would like to thank
  J.-P. Derendinger to draw his attention to this work.} \cite{cornwall74}, see also
  \cite{peskin84,burgess95}. In order to avoid misunderstandings we shortly want to comment
  the relevance of this work for our construction: To get a perturbative
  approximation to the quark-potential a bi-local source
  $\intd{\diff{^4x}\diff{^4y}} \bar{q}(x) K(x,y) q(y)$ is introduced and the
  effective action in presence of this source is calculated directly. As
  result a local (chiral symmetry breaking) minimum is found, but the effective
  action is not bounded from below, but falls off to $- \infty$ as the dual
  field to the source is going to infinity. Using this effective action in our
  recipe for finding the physical minimum would lead either to the
  conclusion, that the situation is unstable or that our procedure is not
  applicable. But this is incorrect: Our procedure insists on the QEA being
  the Legendre transformed of the energy functional $W[J]$ as given in
  equation \eqref{eq:effact} and thus being
  convex. Direct calculations of the QEA need not lead to convex functions:
  Several minima can occur and the function needs not be bounded from
  below. Before such a QEA can be used in our procedure the convex shell has
  to be taken, removing in our case the instability. Indeed the local minimum
  of the QEA by Cornwall, Jackiw and Tomboulis is physical while the
  instability stems from the non-locality of the source \cite{banks76}.
\item Within the restrictions already discussed the QEA should then be well
  defined for finite $m$. The defining fields are the classical fields of the
  Lagrangian multiplet $\tilde{J} \sim \lomega \inv{8 \cg} \tr W^\alpha
  W_\alpha \romega$ and we assume that superspace can be reconstructed on
  these three components at least for the local part in the static limit. To
  distinguish this superfield from the set of its components we will refer to
  it as $\Phi$:
\begin{align}
\label{eq:defPhi}
  \Phi &= \varphi + \theta \psi + \theta^2 L & \varphi &\sim \lomega \lambda \lambda \romega & \lomega \mathcal{L}
   \romega &\sim \tau L + \hc 
\end{align}
  $N=2$ SYM shows explicitly that the local part of the QEA derived this way
  is unacceptable as \emph{dynamical} result \cite{bergamin00:1}. 

Although
  the theory can now be formulated using the dual fields, only, it is useful
  to re-introduce some sources as discussed in section \ref{sec:nonpft}. This
  is specifically done for the sources breaking the symmetries in question, as
  the trigger term is the constant source going to
  zero in the relaxing limit, while the value of the spontaneous parameter is
  an unknown quantity. For $N=1$ SYM this trigger
is the gluino mass and thus $\Gamma[\tilde{J}, \Tilde{\Bar{J}}]$ is replaced
by $\Gamma[\tilde{J}, \Tilde{\Bar{J}}; m_0, \bar{m}_0]$ where $m_0$ is the
constant part of the source $m$. As the mapping of the dual field onto its source must
  be one-to-one in the region where the effective action is well defined, we
  can freely replace the dependence on the dual field by a dependence on the
  re-introduced source. But this dependent variable is not a function of its
  dual field, only, but can depend on all dual fields even in a
  non-holomorphic way (remember that it plays the role of a coupling constant;
  renormalization of the latter need not respect holomorphy as our source
  extension is not restricted by a non-renormalization theorem). In contrast to the LEEA of equation \eqref{eq:geomefflag} we thus
  neither assume that superspace can be reconstructed on the level of the
  three possible sources $\tau$, $\eta$ and $m$ nor
  do we require a holomorphic dependence of the superpotential on these three
  parameters. Instead all quantities depend parametrically thereon.

  Note that in our concept the effective action is now a function of the
  classical fields (constrained by geometry) and of the static part of the
  sources (having a parametrical dependence and including the YM coupling
  constant $\tau$). But this dependence is defined in terms of a single source
  multiplet. There exists the possibility to define two source multiplets, one
  used in the Legendre transformation and the other one used as independent
  variable. Now the effective action is a functional of local classical fields
  as well as of local sources. This conceptually different ansatz has been
  discussed in \cite{burgess95}.
\item In the limit of vanishing gluino source $m$ our concept of global sources
  is problematic as $\varfrac[W]{\tau(x)}|_{\tau \rightarrow \mbox{\tiny
  constant}} = 0$ is true for any value of $\tau$ if supersymmetry is
  unbroken. This just represents the fact that unbroken supersymmetry for a
  coupling constant $\tau$ means unbroken supersymmetry for $\tau + \delta
  \tau$, too. Therefore we have to relax $\tau$ to its
  quantum-mechanical value before relaxing $m$. But this condition is not new
  as exploring the hysteresis line means that we relax the source which breaks the
  symmetry in question (in our case $m$) in the very end.
\item Besides the ones discussed above other problems of the QEA
  especially dangerous to supersymmetric theories have been brought up (see
  e.g.~\cite{intriligator96}). We can just stress again the following points:
  It is absolutely necessary to allow for explicit supersymmetry breaking
  terms regardless of any unbeloved consequences on the geometry of the
  theory. Moreover we have already pointed out that we should use this
  procedure to find the minima, only. Indeed we are not able to show that some
  candidate for the true ground-state found this way is unique and we can thus
  never expect that our QEA captures the whole dynamics over this ground-state
  correctly.
\end{itemize}
As final remark of this section  we would like to mention the analogy of our proposals to QCD: In
  analytic calculations LEEA's and low energy effective Lagrangians have not
  been successful to determine the vacuum structure of QCD but their success
  relies on the fact that the vacuum is known from experiments. We think that
  this order (first the vacuum then the low energy approximation) is crucial
  for any theory with a non-perturbative sector that is not available for exact
  calculations.


\section{Constraints on Dynamical Supersymmetry Breaking}
We want to leave for a moment the construction principles of our effective
action and discuss some constraints on dynamical supersymmetry breaking
independent of the problems mentioned above. The first point are current
algebra relations that lead to the postulation of a massless goldstino if
supersymmetry is spontaneously broken and give a constraint on the value of
the vacuum-energy. If supersymmetry is unbroken the covariant Hamiltonian and
its expectation value with respect to an arbitrary
state $\ket{\psi}$ and to the ground-state $\romega$ are given by
\begin{align}
  \label{eq:hamilton}
  \mathcal{H} &= \inv{2N} \sum_i \bigl( \{ Q^i_1,\bar{Q}_{1i}\} + \{
  Q^i_2,\bar{Q}_{2i}\}\bigr) & \bra{\psi} \mathcal{H} \ket{\psi} &\geq 0 & \lomega
  \mathcal{H} \romega &= 0
\end{align}
where $N$ is the number of supersymmetries and
$Q^i_\alpha$ is the supercharge of the $i$-th supersymmetry. If supersymmetry
is broken the super-charges are no longer well defined. For a single
supersymmetry with supersymmetry-current $S_\mu$ the
local version of the above relation leads to the famous order parameter of
supersymmetry breaking \cite{salam74}:
\begin{align}
  \label{eq:emtensor}
  \intd{\diff{^4x}} \partial^\mu \lomega \mbox{T} S_{\mu \alpha} (x)
  \bar{S}_{\nu \dot{\beta}} (0) \romega &= 2 \sigma^\rho_{\alpha \dot{\beta}}
  \lomega T_{\nu \rho} \romega = 2 \sigma^\rho_{\alpha \dot{\beta}}\, \epsilon_0
\end{align}
and $\epsilon_0=0$ means unbroken supersymmetry, $\epsilon_0 > 0$ spontaneously
broken supersymmetry while $\epsilon_0<0$ would signal a supersymmetry
anomaly. Unfortunately the (perturbative) quantization of  gauge theories destroys
equations \eqref{eq:hamilton} and \eqref{eq:emtensor}: The supercharge of the
quantized theory is not time-independent and the Hamiltonian is not
expressible in the form \eqref{eq:hamilton} \cite{rupp01,rupp01:2}. A time-independent
charge is found after projecting onto the physical Hilbert-space, only. At the
moment we are not able to decide whether the positivity property of
$\epsilon_0$ survives the perturbative quantization or not. Greens functions
with one or more insertions of the supercurrent have been studied recently
\cite{erdmenger98:1,erdmenger98:2,erdmenger99} but the verification of
constraints on supersymmetry breaking from equation \eqref{eq:emtensor} is not yet
possible\cite{sibold00}. This uncertainty relativizes all standard arguments
about dynamical supersymmetry breaking as well as our discussion. We will
assume in the following that at least after projecting onto the physical
Hilbert space the positivity constraint still holds.  Within the
context of our work this assumption is
certainly justified: If supersymmetry is really unbroken in perturbation
theory, the fundamental relations of its algebra must be realized at least on
the physical Hilbert space. If this were not the case, a completely new
understanding of supersymmetry would be necessary. Moreover we follow the standard assumption that the unphysical fields
introduced by the quantization do not contribute to the spontaneous parameters,
i.e.~operators including them have vanishing vev's. For SYM equation
\eqref{eq:emtensor} together with the trace anomaly then leads to
\begin{align}
  \label{eq:SYMorder}
  \lomega {T^\mu}_\mu \romega &= - \frac{\beta}{g} \lomega \mathcal{L} \romega = 4\,
\epsilon_0 \geq 0
\end{align}
and the vev of the Lagrangian becomes the order parameter of supersymmetry
breaking. The fact that the vev of  the Lagrangian must be
positive to enable supersymmetry breaking is a severe constraint on the
spontaneous parameters of this theory. As a side-remark we want to note that
in our approach supersymmetry cannot be broken directly by a gluino
condensate as the latter is the lowest component in the defining superfield.
\subsection{Supersymmetry and the Sign of $\mathbf{\lomega F_{\mu \nu} F^{\mu
      \nu}\romega}$}
\label{sec:fmunusign}
Following our assumption that explicitly broken SYM has a similar vacuum
structure as QCD the number of spontaneous parameters seems to reduce to
$\lomega \tr F_{\mu \nu} F^{\mu
      \nu}\romega$ and $\lomega \tr \lambda \lambda \romega$. The remaining
operators in the Lagrangian should have vanishing vev's and by the assumption
of a smooth dependence on $m$ this should hold at the supersymmetric point,
too. Thus equation \eqref{eq:SYMorder} reads $\lomega \tr F_{\mu \nu} F^{\mu
      \nu}\romega \leq 0$, which is a remarkable result. Completely
  independent of supersymmetry we can ask whether there exists a constraint on
  the sign of $F^2$ and all arguments suggest the same result: $\lomega \tr F_{\mu \nu} F^{\mu
      \nu}\romega \geq 0$ and supersymmetry breaking seems to be excluded as
  the trivial result $\lomega F^2 \romega = 0$ remains, only. For completeness we would like
  to list some of the arguments:
\begin{description}
\item[Sum rules] Based on the work by Shifman, Vainshtein and Zakharov
  \cite{shifman79:1,shifman79:2} the value of $\lomega \itindex{(F^2)}{QCD}
  \romega$ has been estimated to be about $0.250 \mbox{GeV}^4$ (see e.g.\
  \cite{narison98} for recent results on this topic).
\item[Non-decoupling theorem] If the theory depends on the gluino mass
  smoothly we can study the limit $m \rightarrow \infty$. Indeed the trace
  anomaly leads to an interesting relation ($\mathcal{L}$ still represents the SYM
Lagrangian of equation \eqref{eq:classlag}):
  \begin{equation}
    \begin{split}
    {T^\mu}_\mu &=  - \frac{\beta}{g} \mathcal{L} + (\frac{m}{2 \cg} \tr
    \lambda \lambda + \hc )\medsp
    \beta(g) &= - \itindex{\beta}{YM} (g) + \beta_\lambda (g)
    \mbox{\hspace{2.5cm}} \itindex{\beta}{YM} > 0 \mbox{\hspace{0.25cm};\hspace{0.25cm}}
    \beta_\lambda > 0
    \end{split}
  \end{equation}
  Imposing the constraint that in the limit $m \rightarrow \infty$ the trace
  anomaly reduces to the known result of pure gluon-dynamics and taking the
  vacuum expectation value we get:
  \begin{equation}
    \frac{\beta_\lambda(g)}{4 g^2 \cg} \lomega \tr F_{\mu \nu} F^{\mu
    \nu}\romega = - \lim_{m \rightarrow \infty} (\frac{m}{2 \cg} \lomega \tr
    \lambda \lambda\romega + \hc )
  \end{equation}
  Of course this relation is only meaningful if SYM indeed
  tends towards gluon-dynamics in this limit. There is in fact a simple constraint on this relation
  stemming from the vacuum angle: Thermodynamical restoration of CP violation
  \cite{minkowski78:1,minkowski78:2,minkowski90,bergamin00:1} leads in SYM
  with a gluino mass to the following constraints\cite{bergamin00:1}:
  \begin{align}
    (\vartheta - \vartheta_V) + \arg m &= 0 & m \lomega \tr \lambda \lambda
    \romega &= \bar{m} \lomega \tr \bar{\lambda} \bar{\lambda} \romega
  \end{align}
  The fact that the resulting gluon-dynamics must have $(\vartheta -
  \vartheta_V) = 0$ tells us that only real gluino masses can lead to smooth
  decoupling, else the vacuum angle $\vartheta_V$ makes a jump. From the second
  relation we see that in this case the condensate must be
  real. In the limit of a heavy mass the expectation value of $F^2$
  has thus the opposite sign of the expectation value of the gluino
  condensate. The latter sign is negative in analogy to QCD (this already
  follows from PCAC analysis \cite{gell-mann68}, for a
  discussion within QCD see e.g.~\cite{gasser82,leutwyler92}). The notion of
  decoupling a particle by making its mass heavy is
  intuitively pleasing, but it is of course very difficult to make exact
  statements about the behavior of the remaining degrees of freedom. Comparing the
  situation again
  with QCD the non-perturbative region could be crucially different in the latter
  case: While in QCD fractional winding numbers are excluded, they are not in
  pure gluon-dynamics (YM-theory in the following) as well as in SYM. The relevance of
  fractional winding numbers to non-perturbative effects is a highly
  non-trivial problem. It is beyond the scope of this paper to discuss this problem in detail, but
the example again illustrates the importance of BC's. If fractional winding
numbers are assumed to be relevant (see e.g.\ \cite{leutwyler92}) a smooth
decoupling of QCD towards YM is endangered. In contrast to QCD SYM still
decouples smoothly to YM and regardless of this important difference we have
to assume that SYM with a gluino mass has a similar behavior as QCD. A
different point of view has been discussed in \cite{minkowski90:2}. Therein it
is argued that fractional winding numbers are irrelevant in QCD as well as in
YM theory. Then the above problems disappear at the price of a new problem at
the other end of the mass-scale: Now the ground-state of SYM seems to be
degenerate \cite{witten82} leading to domain-walls
\cite{dvali97,kovner97:2}. This degeneracy is an effect of the specific choice
of IR regularization and disappears with fractional winding numbers
\cite{leutwyler92}. In the context of integer winding numbers the degeneracy
is found to be lifted by a more involved study of the thermodynamical limit, as
the phase of the condensate is getting fixed by our program discussed in
section \ref{sec:nonpft} \cite{bergamin00:1,bergamin01:2}.
\item[Euclidian Background fields] Stability conditions on constant gauge fields have
  been studied in \cite{leutwyler80,leutwyler81} and its significance as
  semi-classical ansatz for the YM vacuum has been discussed in
  \cite{minkowski81,efimov98}. These authors study the heat-kernel of
  Yang-Mills theories and therefore the constraints have to be understood in
  Euclidian space. Nevertheless it is worth mentioning the agreement of these
  results: Field configurations are stable if $\lomega \mathbf{E}^2_M \romega
  = - \lomega \mathbf{E}^2_E \romega \leq 0$.
\item[Minkowskian Background fields] The study of Minkowskian background
  fields in gauge theories goes back to the work of Euler/Heisenberg
  \cite{euler36} and Schwinger \cite{schwinger51} on
  QED that led to an important result: If $F^2 < 0$ the potential is not only
  away from its minimum but it is unstable, i.e.~the effective potential
  becomes complex. The generalization of this analysis to YM theories and QCD
  has been performed by Cox and Yildiz \cite{cox80:1,cox80:2}. Although
  non-Abelian gauge theories are much more complicated than QED we expect a complex
  effective potential for $F^2 < 0$ in the
  first case, too.
  \end{description}


\section{Breaking Supersymmetry with $\mathbf{\lomega F_{\mu \nu} F^{\mu
\nu}\romega > 0}$}
\label{sec:auxbreaking}
Can we conclude that either supersymmetry is unbroken or that at least for
small $m$ the vacuum structure is not similar to QCD? We think that this
conclusion is unwarranted. On the level of the field content there exists an
important difference between QCD and SYM: The existence of auxiliary
fields. They play an important role in breaking mechanisms of supersymmetry.
\subsection{The Lagrangian as Auxiliary Field and the Limits of the
  Geometrical Approach}
In the geometrical approach to the effective action there exist two different
types of auxiliary fields: The auxiliary field of the classical field
describing the effective action and the auxiliary field of the underlying
quantum theory. We will refer to them as \second- and \first-generation auxiliary fields
respectively. In the construction of \cite{veneziano82,bib:mamink} the local
part of the 
effective action, being expressed in terms of the Lagrangian- or
anomaly-superfield $\Phi$ (see equation \eqref{eq:defPhi}), is of the form
\begin{equation}
\label{eq:effLag}
 \Gamma[\Phi,\bar{\Phi};m,\bar{m}] = - \intd{\diff{^4 \theta}}
 K(\Phi,\bar{\Phi};m,\bar{m}) + \bigl(\intd{\diff{^2 \theta}} W(\Phi;m,\bar{m}) + \hc\bigr)
\end{equation}
The effective potential then reduces to \cite{bib:mamink}
\begin{align}
\label{eq:effPot}
  \itindex{V}{eff} &= \inv{V^4}\bigl( - \bar{L} \metr L + (L \wphi +
  \hc)\bigr) 
\end{align}
If \eqref{eq:effLag} should represent a meaningful Lagrangian in an expansion
up to second order derivatives as in \cite{veneziano82}, $\metr > 0$ and the
potential as a function of $L$ is not bounded from below and does not even
have a local minimum. Looking at the point
$m=0$ only, this is not surprising: $L$ is the auxiliary field
which has a definite interpretation within supersymmetry. The potential is
getting maximized with respect to $L$ and the remaining (physical) potential is positive
semi-definite. The concave
potential of the  auxiliary field is harmless as long as the full effective action has no
derivative-terms acting thereon: Its equations of motion are algebraic and
the negative semi-definite potential does not have a physical meaning in the
sense of our discussion in section \ref{sec:nonpft}. In contrast to a
different interpretation of the auxiliary field introduced below we call this
behavior non-dynamical. When studying a dependence on $m$ however this behavior is
particularly dangerous: Our extension of the system has been arranged in a
supersymmetry covariant way for any finite $m$. In a naive application
the above structure would be true even for large $m$ and pure gluon-dynamics
would have a reasonable approximation in terms of an auxiliary
field, i.e.\ its low energy approximation would not
have any derivative terms at all. The ansatz would then be  wrong for large $m$ and thus for any
finite $m$ and according to our discussion it would be useless for studying
supersymmetry breaking. Of course such a criticism of the work by Veneziano
and Yankielowicz is --as it
stands-- not acceptable: The effective Lagrangian has been arranged for
vanishing gluino mass and certainly a naive extrapolation to finite masses
does not capture all dynamical effects that could take place in such a deformation. In the remainder of this section we want
to argue that even a more careful treatment must lead to a similar conclusion.

We will focus on the possibility of unbroken supersymmetry at $m=0$ which is
the only scenario compatible with \eqref{eq:effPot}. At $m=0$ thus $\wphi = 0$
and $\wphi<0$ is possible for $m \neq 0$ leading to an acceptable vev of
$F^2$. If
\eqref{eq:effLag} shall represent the full dynamics of the system our
conclusion is certainly correct: \eqref{eq:effLag} has a positive
semi-definite convex potential in the physical fields for all values of $m$
and the number of derivatives is restricted to two. By its construction this
effective action can never break down as the momentum-expansion is always
exact and the potential always stable -- the discussion of the second statement
is analogous to the discussion of this point in the more general model
below.

We thus conclude that \eqref{eq:effLag} does not represent the full effective action
but it is assumed that the effective potential \eqref{eq:effPot} describes at
least qualitatively the correct minima of the theory. This implies
$\metr > 0$, else the potential is either trivial or not bounded from below after eliminating
the auxiliary field. In fact all other interpretations fail to be applicable: A
non-trivial phase of $\metr$ would lead to an unstable potential and with
$\metr < 0$ the potential for the gluino condensate is not bounded from
below. Of course these strict conditions hold in
the minimum, only. Away from the minimum different complex phases may
appear.

\begin{figure}[t]
\begin{center}
    \includegraphics[{scale=0.35}]{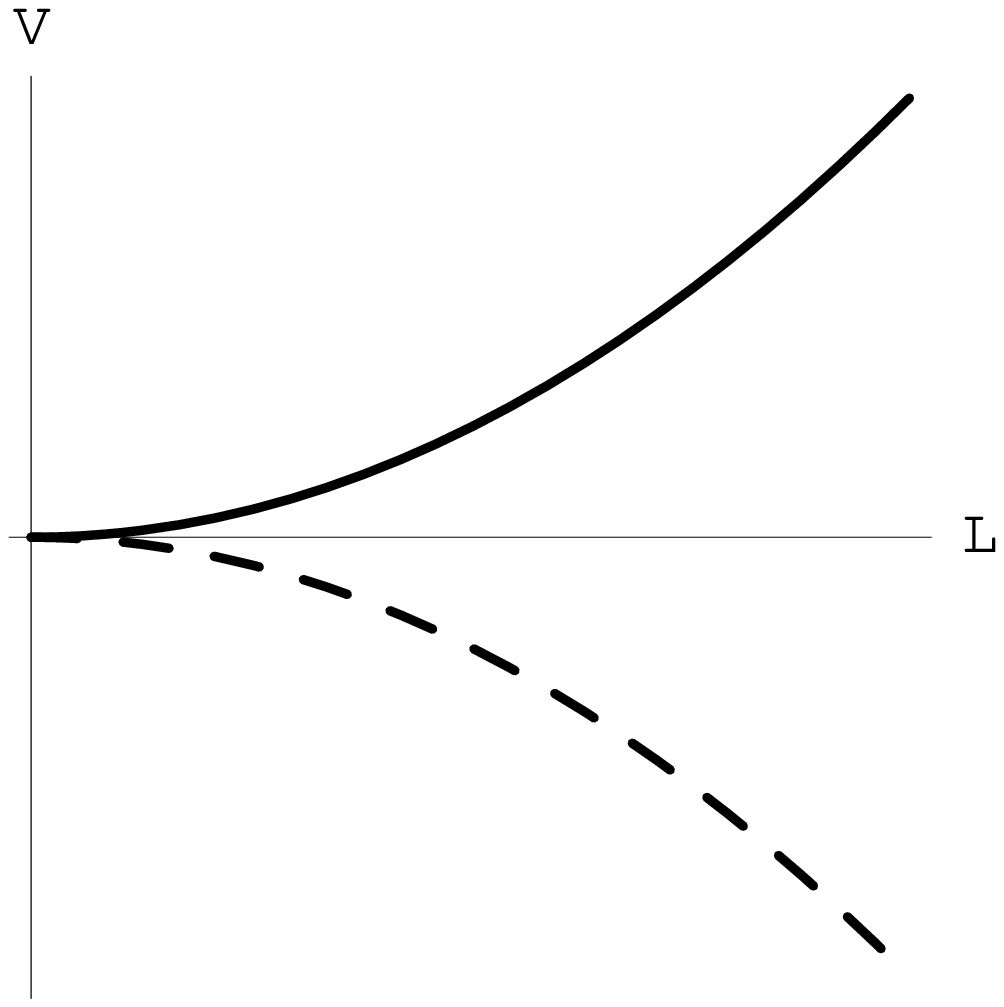} \hspace{3cm}
    \includegraphics[{scale=0.35}]{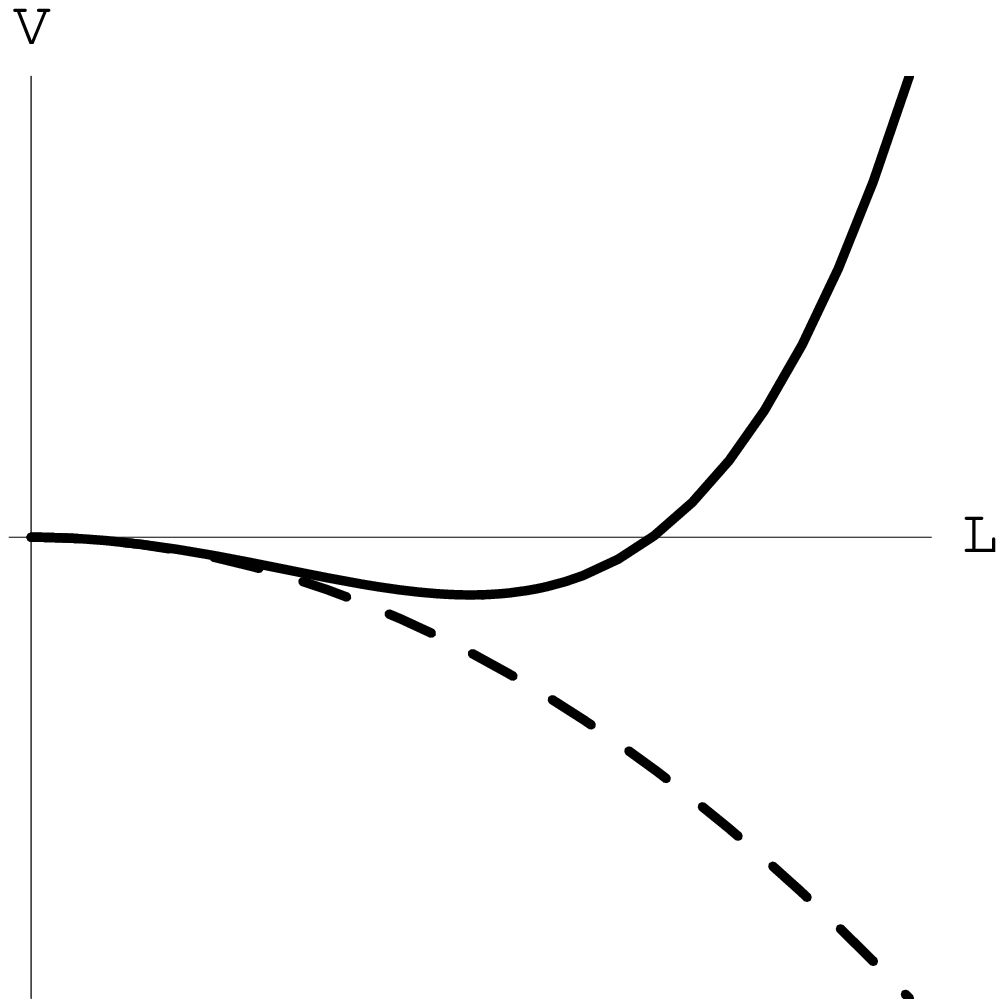}
\end{center}
    \caption{Possible forms of a nontrivial potential of the classical
    Lagrangian. The new contributions can turn the maximum at  $L = 0$ into a minimum (left hand side) or they
      can move the minimum away from the origin (right hand side). The dashed
      line represents the perturbative potential in $L$, $V = - \bar{L} \metr
      L$, the solid line the complete non-perturbative potential.}
    \label{fig:phasetrans}
\end{figure}

The geometrical effective potential is embedded in a more complex effective action
including derivative terms and additional potential terms\footnote{The
  importance of some dynamical arguments in the following does not stand in
  contradiction to the limited relevance of our effective action. If the
  extremalization of the effective potential leads to a maximum in some field,
  the latter must be non-dynamical if the corresponding state plays any role
  in the true ground-state. Our effective action may be incomplete at $p\neq
  0$ as we may have
  missed some physics not reachable by our extension. But this is not
  important here, as we only need to know that there \emph{are} some
  dynamics.}. We cannot specify
their form but only some conditions: At $m=0$ $L$ is an auxiliary field, else
there are dominant contributions to the effective potential not included in
\eqref{eq:effPot}. At $m \rightarrow \infty$ $L$ must become a dynamical field
and the potential must have a minimum in $L$ with $L_0 < 0$. This change of the
behavior of $L$ implies the existence of a phase transition: The effective
potential is always in its allowed region, i.e.~we certainly have a real
$\itindex{V}{eff}$ for all $m$ defined over the range $0 \geq L \geq -
\infty$. Whatever the (static or dynamical) part of the effective action is
doing between $m = 0$ and $m=\infty$, if it wants to turn $L$ from an auxiliary
field into a dynamical field, the potential must at some point be completely
flat. Even if this is thought to be a too strong conclusion in the given
approximation the following points are certainly true: The potential is at
some point zero at $L = \infty$ and there exists a region where it is (almost)
flat around the maximum (turning into a minimum). This is completely sufficient to see that
the system would be unstable. Thus we conclude that there exists a phase
transition at some critical
value of the gluino mass $m_c$. There are two qualitatively different ways how such a phase
transition could look like as shown in figure \ref{fig:phasetrans} (remember
the constraints $L \geq 0$ and $\im L = 0$). The new
contributions to the potential above the phase transition can just turn the
maximum into a minimum leading to a formally smooth value of $L$
in the whole range of $m$ or the value can make a jump at the phase
transition. Despite the fact that $L$ is smooth in the first
case some other parameter has to make a jump, the example just shows that
$L$ need not be the order parameter distinguishing the two
different phases.

We stress that this conclusion is correct even if our effective
action does not represent the theory for all gluino masses $m$ (which will be
important in section \ref{sec:fundaux}). It could
happen that the set of relevant classical operators is different for different
regions of the gluino mass. But the question whether $F^2$ is dynamical or not
is a problem of physics that must be represented correctly in all possible
QEA's. Thus our conclusion is correct if $F^2$ is a relevant low energy degree
of freedom for all $m$, which is included in our assumption of a QCD-like behavior.

Using again the comparison with QCD we expect the phase transition at
$m_c=0$. This possibility indeed exists in the analysis of
\cite{bib:mamink} and cannot be excluded in our discussion. We just would
like to stress the consequences thereof: The phase transition is associated
with the spontaneous parameter of $F^2$, a non-perturbative effect. Such a
phase transition is particularly dangerous to all other non-perturbative
effects, namely chiral symmetry breaking and confinement, which have been
implemented by assumption. In fact the solution in \cite{bib:mamink} suggests
unbroken chiral symmetry which is consistent with \cite{veneziano82}. Besides
these more technical problems the system would be highly unstable and we do
not see how this could still be an acceptable field theory.

In the alternative scenario a phase transition does not exist and additional
contributions to $\itindex{V}{eff}$ are relevant even at $m=0$. In particular
the effective potential does not get maximized with respect to $L$, but
minimized. This does in principle not stand in contradiction to $L$ being an auxiliary
field at $m=0$, but opens new possibilities for supersymmetry breaking.

This discussion shows that indeed the
effective Lagrangian by Veneziano and Yankielowicz suffers from a similar
problem as the LEEA though it does not introduce a IR regulator. The meaning
of a hysteresis curve in the context of this approach remains mysterious and
thus it is not a suitable ansatz in the light of our considerations of section \ref{sec:nonpft}.
\subsection{The Role of the Fundamental Auxiliary Fields\label{sec:fundaux}}
We want to discuss
some consequences of a scenario without phase transition. In this scenario
there appear explicit derivative terms in the auxiliary field at least above
some scale of the gluino mass and thus the \second-generation auxiliary field
$L$ changes its character towards a physical field. Many points of this
section are highly speculative and certainly more investigations are needed to
develop a concrete model where the effects proposed in the following could be
studied.

We have studied the \second-generation auxiliary field without noting the
possible importance of the \first-generation. Indeed more carefully the
constraint derived from equation \eqref{eq:SYMorder} reads:
\begin{equation}
  \label{eq:auxorder}
  \inv{\cg} \lomega \tr(\inv{4} F_{\mu \nu} F^{\mu \nu} - \inv{2} D^2) \romega
  \geq 0
\end{equation}
and supersymmetry is broken if and only if the auxiliary field gets a
non-perturbative vev with $\lomega D^2 \romega > \inv{2} \lomega F^2 \romega$.
The fact that supersymmetry breaking is driven by the vev of the auxiliary
fields is an old wisdom from perturbation theory
\cite{raifeartaigh75,fayet74}. As many restrictions on perturbative supersymmetry (by
assumption) hold in the non-perturbative region, too, the importance of the
auxiliary fields therein is not surprising. Breaking supersymmetry by
postulating non-trivial dynamics of the auxiliary fields does certainly not
look very appealing, but under the given assumptions it is a correct and
necessary proposition. We want to point out some restrictions and consequences
of this scenario:

Non-trivial dynamics of the auxiliary field lead to a complete breakdown of
the supersymmetry covariant approach: As the auxiliary field changes its
character towards a physical field, the extremalization of the potential must
lead to a minimum and non-geometrical contributions are relevant. Moreover all
supersymmetry covariant expressions (sources and BC's) depend on the
combination $L = - \inv{4} F^2 + \half{1} D^2$, only. Of course we can
determine the minimum $\delta_{L} \itindex{V}{eff}(L_0,\bar{L}_0;m,\bar{m}) =
0$. As $L_0$ is directly related to the goldstino coupling the covariant
effective action has still a meaning at least for small masses $m$ where the
\mbox{(pseudo-)} goldstino is a special particle. Above this scale the
combination looses its meaning and physics are probably not described by these
combinations any more. In contrast to the case with $\lomega D^2 \romega = 0$
trivially however, the goldstino coupling is not a primary object but we have
to study $F^2$ and $D^2$ independently. Trying to impose a constraint on
supersymmetry covariant objects only, leads to difficulties: Infinitely many
combinations of the gluon- and the auxiliary-field lead to a specific value of
$L$ (even for $L=0$). In the IR regularization or in a semi-classical
calculation this leads to a summation over all these combinations and the vev
of a single operator $F^2$ or $D^2$ is no longer well-defined. By treating the
two operators as independent objects $L(m)$ (or the dependence of
$L$ on any other external parameter) describes a line in the $F^2$-$D^2$ plane. Unbroken supersymmetry would imply
that the line starts at the origin (supersymmetric point) and that $F^2$
develops a vev as $m$ increases. For broken supersymmetry the shape of the
line is unknown. It starts at some point with $2 D^2 > F^2$ and this
constraint is fulfilled within the range of the pseudo-goldstino being a
special particle. Above this range the combination $L$ is no longer meaningful
and thus the line is unimportant (or perhaps not even defined). Besides $L(m)$
which could be calculated for small $m$ by a chiral perturbation theory for
the goldstino and in the large $m$ limit by using YM-results, independent
knowledge about one of the two involved basic operators is needed. Finding
this line would answer many open questions about dynamical symmetry breaking
in SYM and must be one of the main areas of future research.

In this scenario supersymmetry breaking is a non-perturbative
non-semiclassical effect: It can be established from an IR-regularization
mixing the physical fields with the auxiliary field, only (i.e. the separation
of the path integral into a physical and an auxiliary field part is no longer possible).
Clearly spaces allowing the definition of instantons are not sensitive to
non-perturbative effects of $D^2$. Instanton calculations have been performed
in different regions \cite{amati88,shifman88,kovner97} and have found to be
consistent with each other within the semiclassical approximation
\cite{hollowood99,davies99}. In agreement with our discussion supersymmetry
does not break by instanton induced effects.

Which are the spaces that make effects from auxiliary field visible? A simple
analysis of the above separation condition shows that source extensions alone are
useless. Some sources like the goldstino source couple the auxiliary
field to the physical ones. But neither do they lead to derivative terms in
the auxiliary fields nor do they change the sign of the
potential, as this requires effects from non-renormalizable operators. There
is room for speculations within more general BC's, as non-renormalizable operators
can now be included, but at the moment we are not able to suggest any
concrete calculation that could test our proposition.

Do the auxiliary fields turn into physical fields completely or are they still
non-dynamical in the end? If $\lomega D^2 \romega \neq 0$ this can only be due to quantum
fluctuations and there must be a finite correlation
$\lomega D(x) U(x,y) D(y) \romega \neq 0$ at least for small distances. This all happens due
to an infrared effect and thus at least in this region the auxiliary field is
indeed a physical field. Two different interpretations are possible: The \first-generation auxiliary field is non-dynamical
regardless of the value of $m$ after removing all IR regularizations. Thus we catch all its important effects by
replacing $D^2$ by its vev in the classical Lagrangian. This leads to an
alternative interpretation of this constant: Hughes and Polchinski
\cite{hughes86:2} have shown that equation \eqref{eq:emtensor} can
consistently be generalized to
\begin{align}
  \label{eq:emtensor2}
  \intd{\diff{^4x}} \partial^\mu \lomega \mbox{T} S_{\mu \alpha} (x)
  \bar{S}_{\nu \dot{\beta}} (0) \romega &= 2 \sigma^\rho_{\alpha \dot{\beta}}
  \lomega T_{\nu \rho} \romega + C
\end{align}
where $C$ is a dynamical parameter and exactly represents the vev of the
fundamental auxiliary field after its elimination.

In the second interpretation the \first-generation auxiliary field remains dynamical in the
thermodynamical limit. At the moment this is pure speculation and we cannot
give any similar model where this would happen.

Two important points in our discussion are highly speculative:
\begin{itemize}
\item The first point are the new contributions to the effective action, which
  are not of the form \eqref{eq:geomefflag} changing the potential of
  the \second-generation auxiliary field and leading to derivative terms
  thereof. The existence of such contributions immediately raises the
  questions of the realization of supersymmetry on the level of the QEA and of
  the validity of a momentum-expansion of the latter. It is beyond the scope
  of this paper to answer these questions here. A simple model of this type will be presented elsewhere
  \cite{bergamin02:1}. We also want to note that within a quite different
context the effect of turning an auxiliary field into a dynamical one is known:
In effective actions of SQCD based on gauged non-linear sigma models
\cite{nemeschansky85,bergshoeff85,karlhede87,gates97}. We should be careful in
deriving any conclusions from this but the effect itself shows that the
application of superspace geometry is not at all straightforward.
\item Moreover it is not understood, which role the stability constraint
  from section \ref{sec:fmunusign} plays from the point of view of the
  fundamental theory. All arguments in this paper have been semiclassical and
  we do not know exactly to what objects the symbols $\lomega F^2 \romega$ and
  $\lomega D^2 \romega$ refer to. The dynamics and the vev of the
  \first-generation auxiliary field discussed in this section apply to the
  semiclassical object $\lomega D^2 \romega$ and at the moment we are not able
  to conclusively relate this object to any known characteristic of the
  underlying theory.
\end{itemize}
Finally we
want to point out that the assumption of vanishing vev's in the ghost sector
does not contradict to our proposition of a non-trivial $D^2$: The quantization
of gauge theories can be performed in many different ways and depending on the
procedure different unphysical fields appear. The existence of the auxiliary
fields in supersymmetry however is unambiguous in the classical and quantized
theory.


\section{Discussion of the $N=1$ Result}
Before going into the discussion of our results we want to summarize the four
different low energy behaviors of SYM that we found:
\begin{enumerate}
\item Supersymmetry is unbroken.
  \begin{enumerate}
  \item \label{itemoa}Neither \first\ nor \second\ generation auxiliary field receive
    non-perturbative contributions. This implies the existence if a phase
    transition in the variation of the gluino mass with $m_c = 0$. We expect
    that all condensates vanish and conclude that the theory does not have an
    acceptable infrared behavior.
    \item \label{itemob}The \first\ generation auxiliary field can be eliminated consistently
      but the potential in $L$ has a minimum due to
      non-perturbative contributions. For $m=0$ the minimum must be at
      $L = 0$ and supersymmetry is unbroken.
    \item \label{itemoc}Neither \first\ nor \second\ generation auxiliary field
      behave non-trivially, but the minimum for $m=0$ is still at the supersymmetry
      conserving point.
  \end{enumerate}
\item \label{itemt}Both auxiliary fields get non-perturbative contributions
  and supersymmetry breaks dynamically.
\end{enumerate}
For \ref{itemoa} the structure of the vacuum has been discussed. Especially in
the cases \ref{itemob} and \ref{itemoc} the latter is unknown as the
geometrical approach for effective Lagrangians does not lead to the correct
results. From our point of view it would be very surprising if the auxiliary
fields could get non-trivial contributions without breaking
supersymmetry. Provided $N=1$ SYM does exist as a quantum theory and can be
described at low energies by the effective action defined in this work the favorite
for its low energy behavior is the scenario \ref{itemt}.
\subsection{Dynamical Supersymmetry Breaking and the Witten Index}
When we want to give an interpretation of our discussion two questions arise:
Although the LEEA or low energy effective Lagrangian approaches have serious
conceptual problems, the result derived therefrom could anyway happen to be
correct. Can we exclude this? Besides this semiclassical approximation other
arguments for unbroken supersymmetry have been given. What is their relevance
within our discussion?

One part of the first question has already been answered in the last section:
We cannot exclude a phase transition at $m=0$. The resulting theory has
$\lomega F^2\romega =
0$ and the analysis of \cite{bib:mamink} suggests $\lomega\lambda\lambda\romega = 0$ as
well (this solution stands in agreement with \cite{veneziano82}, further
discussions of this state have been given in \cite{kovner97}). As all
spontaneous parameters vanish we expect that the theory is not
confined either. Besides the instability to perturbations we do
not expect the IR problem getting solved. Apart from this both solutions
(\cite{veneziano82} and \cite{bib:mamink}) are probably
incomplete as they do not have the correct analytical structure: In
\cite{bib:mamink} $F^2 < 0$ everywhere except at the origin,
\cite{veneziano82} has $F^2 < 0$ in the region between the chirally symmetric
and the chirally broken minimum. None of the two generates a complex phase
within these regions and the strict constraints of the geometry make it difficult
to include this instability (notice that the metric $\metr$ must by its
construction be independent of $L$).

Besides the LEEA we have discussed the instanton calculations and we do not
want to go more into details. An important argument against
supersymmetry breaking seems to be the Witten index
\cite{witten82}. We want to give a brief comment on the work by Witten. Three
ingredients are crucial to come to the conclusion that supersymmetry must be
unbroken: Holomorphic dependence, independence of the boundary conditions and
availability of perturbation theory. Due to the holomorphic dependence of the superpotential
we only need to answer the question of supersymmetry breaking for one
(non-vanishing) value of the coupling constant (the latter is of course chosen
to be small). Now Witten explicitly states that the index of an operator (and
thus the fundamental characteristics of a trial vacuum state) is independent
of the boundary conditions and that he thus is allowed to choose them
arbitrarily. In this general form we do not agree herewith (see
section \ref{sec:nonpft}). After choosing boundary conditions that at least
allow to do perturbation theory of the models in question, Witten argues that
this is enough to calculate the index. To decide whether this simplification
is allowed or not we have to clarify the meaning of the phrase ``a theory in its perturbative
region'': It can mean that perturbation theory gives a reasonable approximation
to the true result, although the series need not converge, or it can mean that
the series really converges. An example of the first kind of understanding is
the vacuum of a non-Abelian gauge theory with small coupling constant. There
are non-perturbative effects (chiral symmetry breaking, dynamical mass gap),
but these effects are very small and thus the non-convergent perturbative
expansion can nevertheless give a good approximation. It is often speculated
that QCD in the de-confined phase is an example of the second kind of
understanding. Indeed all known effects of non-perturbative dynamics seem to
vanish and it could be that the perturbative expansion now converges. Which
kind of interpretation should we use if we want to show that supersymmetry is
unbroken? Certainly the first one. Supersymmetry is unbroken if and only if
the vacuum energy is \emph{exactly} zero. For small coupling
constants a (non-perturbative) supersymmetry breaking effect may be suppressed
exponentially, this does not help as the holomorphy argument is immediately
useless if the effect does not vanish completely. Certainly we cannot show
that the perturbation series for supersymmetry with small coupling constant in
a finite volume really converges.

Can the special structure of Witten's calculation, using a version of the
Atiyah-Singer index theorem, justify a perturbative calculation? An argument
has been given in \cite{witten82} that this could be true: Consider an
operator with non-zero index (say $n \neq 0$) in perturbation theory. If the index shall become
zero non-perturbatively we cannot argue that the $n$ states move away from
zero a little bit: Supersymmetry only allows pairs of states with non-zero energy and thus
the index remains $n$. Non-perturbative corrections therefore have to change the
spectrum completely. Witten argues that such a correction would change the
asymptotic behavior of the potential and he assumes that non-perturbative
contributions cannot do this (at least for small coupling constants). We do not agree with
this conclusion. The effect of non-perturbative
contributions is exactly to change the potential in such a way: Confinement
indeed changes the spectrum of the theory completely and we expect that this
could also change the asymptotic behavior of the potential. Within the whole
discussion the actual
value of the coupling constant is completely irrelevant (as long as it is
non-zero). This difference between the asymptotic behavior of the
perturbative and the non-perturbative potential (for fixed value of the
coupling constant) does not at all stand in contradiction to the fact that the
non-perturbative potential is not allowed to change its asymptotic behavior
when varying the coupling constant. These are two completely independent
properties of the (perturbative and non-perturbative) potential. For
additional discussions of related problems in a semiclassical
analysis we refer the reader to \cite{minkowski81}.

Even if the Witten index would be a correct analysis of the theory within the
special choice of boundary conditions, its consistent interpretation within our framework is quite
easy: The author uses BC's that do not break supersymmetry. Thus we cannot
expect that he will find a supersymmetry breaking state at large
volume. Assuming that the state found in the limit is a reasonable candidate
for the ground-state, is there any argument that it must be the
true ground-state? We have seen that this is not true in general and --in
contrast to a common misunderstanding of constraints from supersymmetry-- the latter
does not help in this situation: The supersymmetric trial vacuum minimizes the
vev of the energy-momentum tensor but (as a function of the classical fields)
this is certainly not the correct quantity getting minimized by the true
ground state (on the semi-classical level this has been discussed for YM  in
\cite{minkowski81}). The analogy of the energy-momentum tensor and the effective
potential holds in perturbation theory due to the non-renormalization theorem,
but the latter need not be valid in the non-perturbative region: Denoting by $\varphi_0$ the value of the fields at
the minimum of the effective potential, the consequences of equation
\eqref{eq:emtensor} are
\begin{subequations}
\begin{align}
  \lomega {T^\mu}_\mu \romega &\geq 0 &&\Rightarrow &
  \itindex{V}{eff}(\varphi_0) &\geq 0 &&\mbox{\parbox{8cm}{In perturbation
  theory by means of the non-renormalization theorems.}}\medsp
  \label{eq:nonpsb}\lomega {T^\mu}_\mu \romega &> 0 &&\Leftrightarrow &
  \itindex{V}{eff}(\varphi_0) &< 0 &&\mbox{\parbox{8cm}{Possible scenario of
  non-perturbative supersymmetry breaking.}}
\end{align}
\end{subequations}
The vev of $D^2$ shows in a very simple way how Witten's vacua become irrelevant: The
minimum in the effective potential lies at $L_0 >0$ and $\itindex{V}{eff}
(L_0) < \itindex{V}{eff}(0)$, but clearly
\begin{equation}
\lomega {T^\mu}_\mu (L_0) \romega = - \inv{4} \lomega F_{\mu \nu} F^{\mu \nu}(L_0)
\romega + \half{1} \lomega D^2(L_0) \romega > \lomega {T^\mu}_\mu (0) \romega
\end{equation}
The effect can take place as the wrong sign
from the classical potential remains in the energy-momentum tensor while it is
getting changed in the effective potential by non-semiclassical contributions.

Though we do not agree
with Witten's interpretation of his calculation, the index can nevertheless
unravel interesting properties of dynamical supersymmetry breaking: If the
index of a theory within a certain choice of BC's is found to be non-zero,
supersymmetry can break dynamically if and only if one of the following points
applies:
\begin{itemize}
\item The full non-perturbative potential has a different asymptotic behavior
  than the approximation used to calculate the index.
\item There exist non-perturbative effects that do not respect the
  non-renormalization theorem and that destroy the perturbative equivalence
  between 
  the vacuum energy and the minimum of the effective potential (cf.\ \eqref{eq:nonpsb}).
\end{itemize}
According to our present knowledge both effects are not only non-perturbative
but also non-semiclassical.
\subsection{Different Results from Different Geometry?}
All results of section \ref{sec:auxbreaking} are valid under specific assumptions on the properties of
SYM under quantization, only. There still exists the possibility that the specific
choice of the geometry is wrong and not the
geometrical approach itself.
We want to make some comments about this problem. The strategy of the
approaches discussed in this work is to use solely gauge singlets as classical
fields. Following the philosophy of semi-classical approximations not to
include non-physical fields from the quantization of the theory, the source
extension is then complete and unique. There exists exactly one
superfield invariant under full susy-gauge transformations whose highest
component is a candidate for the classical action (if there would exist any
other superfields, our classical Lagrangian would not be the most general one
obeying all symmetries and we would have to include this superfield in the
action). Moreover there exists exactly one possibility to extend the coupling
constant supersymmetry covariantly to a superfield. Thus we expect that all
other source extensions at least partially break super-gauge invariance. We
can illustrate this on the basis of the simplest generalization of our
extension: The chiral source extension is problematic as the fundamental
structure of the action is not chiral. Thus it would be most natural to
consider the action as an integral over full superspace and to introduce a
full source-multiplet. We call the new multiplets $\Phi_f$ and $J_f$ with
\begin{align}
  (\Phi_f)|_{\theta^2 \bar{\theta}^2} &= L & (J_f)| &= \tau 
\end{align}
If the chiral extension is indeed inconsistent on the level of quantum
operators, the operator superfield associated with $\Phi_f$,
$\Phi_f^{\mbox{\tiny op}}$, enters in some expression in a non-chiral way,
e.g.\ as $\bar{\Phi}_f^{\mbox{\tiny op}} \Phi_f^{\mbox{\tiny op}}$. In
principle this does not yet imply that the effective action cannot be
described by the geometry of a chiral superfield: If all non-chiral
contributions vanish after taking the vacuum expectation value, the effective
action depends on $\Phi = \bar{D}^2 \Phi_f$ only and the chiral geometry is
getting restored with all the problems discussed in this work. If the
non-chiral expressions do not vanish on the level of classical fields all descriptions based on the
chiral extension are essentially incorrect. This especially means that the
true ground-state of SYM theories can only be found by considering
gauge-symmetry breaking sources and the associated operators must have an important
influence on the structure of the ground-state! Besides the fact that
super-gauge symmetry now breaks dynamically, probably no straightforward
interpretation of such a result could be given as the low energy structure of
SYM now could depend on the unphysical components of the prepotential.

Within the context of non-standard geometries we should address another
problem: the representation of the glue-ball. Indeed we have tacitly assumed
that the operator $F^2$ is directly related to the lowest glue-ball state. This
is not at all obvious but emerges from the fact that $F^2$ is the only
purely gluonic, renormalizable and gauge-invariant operator. In principle
there exists a simple way to keep $L$ consistently an auxiliary field: we have to introduce different
operators for the \second\ generation auxiliary field and the glue-ball
operators, respectively. Based on the Lagrangian by Veneziano and Yankielowicz this has been
proposed by G.R.\ Farrar et al.\ \cite{farrar98}. The authors observe that
after splitting the classical Lagrangian into two independent fields for the
real and imaginary part respectively, the fields can be recombined to a new
superfield $U$ called constrained three-form multiplet (introduced in
\cite{gates81} and used to describe a gauge-theory based on a super
three-form). This new superfield contains additional degrees of freedom
identified with the glue-balls, the original superfield is re-found by the
relation $\bar{D}^2 U \sim \Phi$ whereby $U$ is real. The authors then
construct an effective Lagrangian in $U$ that cannot be written in terms of
$\Phi$ and derive the spectrum of the glue-ball states therefrom. From our
point of view the procedure is problematic: The classical fields are not
primary objects but rely on a source-extension of the Lagrangian. The source
extension corresponding to this extended system is not given in
\cite{farrar98}, we suspect that it does not exist at all. From the relation
$\bar{D}^2 U \sim \Phi$ we conclude
\begin{align}
  U^{\mbox{\tiny op}} &\sim \ndup{A}{\alpha}\nddn{W}{\alpha} + X & \bar{D}^2 X
  &= 0 &\Rightarrow&& X&= \bar{D}^{\dot{\alpha}} Y_{\dot{\alpha}} 
\end{align}
with $\nddn{A}{\alpha} = -i (e^V
\nddn{D}{\alpha} e^{-V})$ representing the spinorial connection. It is easy to check that the reality condition is incompatible with the above
structure. If the source extension shall be defined on the basis of the
operator superfield $ U^{\mbox{\tiny op}} \sim
\ndup{A}{\alpha}\nddn{W}{\alpha} + \hc$ all the terms not compatible with the
constrained three-form multiplet would have to vanish when taking the vacuum
expectation value. But with this choice we immediately get into new difficulties:
Obviously the so-called glue-ball states (e.g.\ the lowest component of $U$,
identified with the scalar glue-ball) are not super-gauge invariant and are
moreover not at all gluonic operators but depend on the non-gluonic physical
fields as well as on the non-physical ones. The example shows that it is very
difficult to include glue-ball states which are not associated with $F^2$. Of
course this does not mean that our solution must be correct, but within all
approaches discussed in this work there is simply no way to describe a glue-ball not associated with
the classical field of $F^2$. A different picture could be described
within a model including non-renormalizable operators as low-energy degrees of
freedom, only.

Recently it has been speculated that the
operator $A_\mu A^\mu$ could have a non-trivial meaning in the QCD vacuum
\cite{gubarev00,verschelde01}. The
local version of this operator is (due to gauge-invariance) no candidate for a
glue-ball state, but the integral over it could indeed be relevant. We are not able to give further comments on these
calculations at this point. Even if this would be a promising ansatz to
understand non-perturbative effects in the gluonic sector it would probably be
very difficult to include this into our approach to SYM.

\section{Some Comments on More Complicated Models}
Let us finally make some remarks on more complicated models: Can the inclusion
of matter help? We are not able to give a final answer but would
like to point out some problems: If we consider SQCD with large masses the
only new contribution to the trace anomaly (the quark condensate) has again the wrong sign from the
point of view of supersymmetry. If the masses are small the situation is more
complicated due to possible contributions from scalar vev's. At zero mass the
new problem of a classical moduli space arises. In the LEEA approximation
\cite{davis83,affleck84,affleck85} the
latter is found to be lifted and the low energy structure is again a SYM
theory. In the light of our
discussions this result is again problematic.

We want to analyze $N=2$ SYM a little bit more in detail.  Within $N=2$ superspace the action is given by
\begin{equation}
  \begin{split}
\mathcal{L} &= \intd{\diff{^4 \theta}} \tau  \tr(W^2) + \hc \medsp
&= \inv{\cg} \tr\bigl( \frac{1}{g^2} [ D_\mu , \bar{C}] [D^\mu , C] +
  \frac{i}{g^2}  \lambda^{i \alpha} \sigma^\mu_{\alpha
  \dot{\alpha}} [D_\mu, \bar{\lambda}_i^{\dot{\alpha}}] -  \inv{4 g^2} F_{\mu \nu} F^{\mu \nu} -
  \frac{\vartheta}{32\pi^2}  F_{\mu \nu} \tilde{F}^{\mu \nu}\medsp
  &\quad + \inv{4 g^2} H_{\{ij\}} H^{\{ij\}} +  \frac{1}{g^2} C [C, \bar{C} ] \bar{C} +  \frac{i}{\sqrt{2} g^2} C \{ \bar{\lambda}^i_{\dot{\alpha}} ,
  \bar{\lambda}_i^{\dot{\alpha}} \} -  \frac{i}{\sqrt{2} g^2} \bar{C} \{ \lambda_i^\alpha ,
  \lambda^i_\alpha \} \bigr)
  \end{split}
\end{equation}
with the chiral $N=2$ multiplet (written in $N=2$ Wess-Zumino gauge)
\begin{equation}
  \begin{split}
      W(x, \theta^\alpha_i) &= \sqrt{2} C(x) + \sqrt{2} \theta^\alpha_i \lambda_\alpha^i(x) +
  \theta^{\alpha \beta} v_{\alpha \beta}(x) + \theta_{ij} H^{ij}(x) +
  \vartheta^\alpha_i \chi_\alpha^i(x) + \theta^4 D(x) \medsp
v_{\alpha \beta} &= \half{i} \sigma^{\mu \nu}_{\alpha \beta} F_{\mu \nu}
  \medsp
  \chi^\alpha_i &= i \sqrt{2} (\bar{\sigma}^\mu)^{\dot{\alpha} \alpha} [ D_\mu ,
  \bar{\lambda}_{i \dot{\alpha}} ] + \frac{i}{\sqrt{2}} [ \bar{C} ,
  \lambda^\alpha_i ] \medsp
  D &=  \sqrt{2} \bigl[ D^\mu , [D_\mu , \bar{C} ] \bigr] - \inv{\sqrt{2}} \bigl[ \bar{C} ,
  [\bar{C} , C ] \bigr] - i \{ \bar{\lambda}^i_{\dot{\alpha}} ,
  \bar{\lambda}_i^{\dot{\alpha}} \}
  \end{split}
\end{equation}
When breaking $N=2$
SYM we have two choices: We can break directly both supersymmetries or we can
break the theory down to $N=1$. Let us look at the first case. We define the
source multiplet
\begin{equation}
        J(x) = \tau(x) + \theta^\alpha_i \zeta_\alpha^i - 4 \theta_{ij} m^{ij}(x) + \theta^{\alpha \beta} w_{\alpha \beta} + \vartheta^\alpha_i \kappa_\alpha^i + 4 \theta^4 M^2(x) 
\end{equation}
which has to obey the stability condition \cite{bergamin00:1}
\begin{align}
        \re(\tau) &\geq 0 &  g^2 \rho^2 &\geq |M^2 + g^2 \mu^2|
\end{align}
with $\mu^2 = \tilde{m}_A \tilde{m}_A$, $\rho^2 = \tilde{m}_A
\Bar{\Tilde{m}}_A$; ${m^i}_j = -i \frac{{(\tau_A)^i}_j}{2} \tilde{m}_A$.

Note that the
stability constraint forbids to choose the scalar mass term $M^2$ to be
non-zero while keeping the fermions massless ($m^{ij} = 0$). To
establish the non-decoupling theorem of $N=2 \rightarrow N=0$ we should start
from the trace anomaly of $N=2$: ${T^\mu}_\mu = - \frac{\beta}{g}
\mathcal{L}$. In contrast to $N=1$ the Lagrangian is now much more complicated
and we cannot assume that the field-strength tensor is the only physical
operator receiving a vacuum expectation value. Moreover the
energy-momentum tensor is conserved for vanishing sources only and thus we
cannot expect that the trace anomaly is $\sim \mathcal{L}$ for $J \neq 0$. We
thus have to define a covariantized energy-momentum tensor that is conserved
for any value of the sources \cite{bergamin01:2}. The basic structure of the trace anomaly of this
covariantized e.-m.\ tensor for any value of the
sources must be given by
\begin{equation}
\begin{split}
  {T^\mu}_\mu &= - \frac{\beta}{g} A + \frac{2}{\cg} (M^2 \tr C^2 + \hc) -
  \inv{\cg} (m_{ij} \tr \lambda^i \lambda^j + \hc)\medsp
  &\quad + \frac{4}{\cg} \tr (m_{ij}
  C + \bar{m}_{ij} \bar{C} )^2 + B
\end{split}
\end{equation}
where $A = - \inv{4 \cg} F_{\mu \nu} F^{\mu\nu} + A'$ and $B$ represents any
additional terms independent of the sources. Now we split the
$\beta$ function into
\begin{equation}
  \beta(g) = - \itindex{b}{YM} + b_\lambda + b_C
\end{equation}
Without dropping any terms the non-decoupling theorem then reads (omitting the
ground-state brackets)
\begin{equation}
  \begin{split}
    - \frac{b_\lambda + b_C}{g} A + \frac{\itindex{b}{YM}}{g} A' &= - \lim_{m
    \rightarrow \infty} \bigl( \frac{2}{\cg} (M^2 \tr C^2 + \hc) -
  \inv{\cg} (m_{ij} \tr \lambda^i \lambda^j + \hc)\medsp
  &\quad + \frac{4}{\cg} \tr (m_{ij}
  C + \bar{m}_{ij} \bar{C} )^2 + B \bigr)
  \end{split}
\end{equation}
Here $m \rightarrow \infty$ means any limit degenerating the theory to YM and
respecting the stability constraints. If the fermions and scalars decouple
from each other in the heavy mass limit (i.e.\ the vev's of fermionic operators
do not depend on the scalars and vice versa) we recover the standard
non-decoupling theorem if $\lomega A' \romega = \lomega B \romega = 0$.  Using
a formulation
of the sources in terms of $N=1$ superfields one can see that the non-decoupling theorem  of the
fermions agrees with the one of a single Dirac fermion. A note about the auxiliary
fields: Of course they could have a vev for vanishing sources, but as
discussed for $N=1$ we expect the latter to disappear at some finite scale of
the mass parameters.

More interesting than breaking both supersymmetries turns out to be the
partial supersymmetry breaking. Now only one component of $m^{ij}$ is getting
large, giving a mass to the scalars as well as to one gluino. The trace
anomaly need not be $N=2$ supersymmetric, but it must still respect $N=1$
supersymmetry. We then end up with the following non-decoupling theorem for
the fermions
\begin{equation}
  \frac{b_\psi}{g^3 \cg} \lomega (\inv{4}\tr F_{\mu\nu} F^{\mu\nu} - \half{1}\tr D^2)
  \romega = - \lim_{m \rightarrow \infty} \frac{m}{2 \cg} \lomega \tr \psi \psi \romega
\end{equation}
where $\psi$ is the massive gluino in the $N=1$ language. We argued in this
work that $\lomega (\inv{4}\tr F_{\mu\nu} F^{\mu\nu} - \half{1}\tr D^2)
\romega < 0$ which would imply $\lomega \tr \psi \psi \romega > 0$, a clearly
unacceptable result. To banish again the vacuum expectation value of the
auxiliary field does not really cure the problem. It would imply $\lomega \tr
\psi \psi \romega \rightarrow 0$ for $m \rightarrow \infty$ which does not at
all fit with our expectations. The short analysis shows that our arguments
about $N=2$ have probably been too simple, similar questions may arise when
studying the signs of the scalar condensates. We can
not yet give an acceptable interpretation of this problem but have to postpone
this to future investigations. Of course SQCD will have a similar
behavior when decoupling some quark superfields. 

Let us finally look at $N=2$ SYM near the symmetry conserving point. Is there
need for a vacuum expectation value of the auxiliary fields as in $N=1$? When
reducing the potential to
\begin{equation}
V = V(\phi) = \inv{g^2 \cg} \tr[\phi, \phi\dega]^2
\end{equation}
as done by Seiberg and Witten \cite{seiberg94}  all
contributions to the trace anomaly are negative semi-definite
\begin{equation}
  \lomega {T^\mu}_\mu \romega = - \frac{\beta}{g} \lomega \mathcal{L} \romega
  \rightarrow \frac{\beta}{g^3 \cg} \lomega \tr[\phi, \phi\dega]^2 \romega \leq 0
\end{equation}
and supersymmetry is broken if and only if the auxiliary field changes its
character. Completely analogous to $N=1$ the problem of a phase transition
emerges. The situation is even more delicate than in $N=1$: If the
Seiberg-Witten solution is correct while $N=1$ is broken, $N=2$ needs
not only be protected from YM theory but also from $N=1$ SYM and we end up
with a phase transition in the gluino mass. This phase transition is again
expected at $m=0$ and is associated with a jump in the modulus. If on the other hand
the $N=2$ auxiliary fields  get a vacuum expectation value, the auxiliary field of the
matter field  must be non-trivial at least for small values of the
breaking parameters. This raises the question of the role of these auxiliary
fields that could not be considered in this work. However
the structure of $N=2$ SYM is much more complicated and we thus should be
careful with the relevance of these statements. It only shows that the solution
by Seiberg and Witten suffers from the same problem as $N=1$ SYM.

Considering the special structure of $N=2$ theories a new question arises: It
has been shown by Olive and Witten \cite{witten78} that a theory with
classically vanishing central charges can develop them dynamically leading to
magnetic monopoles. This behavior is an important assumption in the solution
by Seiberg and Witten for $N=2$ SYM and SQCD. The existence of magnetic
monopoles within a full quantum theory (i.e.\ not as semiclassical
approximation) is however unclear. Striebel \cite{striebel87} showed that the
finite energy solutions of the magnetic monopoles solely exist if the
background field of the full gluon sector vanishes. This does not have direct
implications to the solution by Seiberg and Witten. These authors look at the
gauge group $\su(2)$ broken down to $\uone$, only. Then we indeed expect that
$F^2$ vanishes. When choosing a more complicated gauge group however we will
in general end up with partial breaking of gauge symmetry. $F^2 \neq 0$ then
implies the absence of magnetic monopoles and the dynamical generation of
central charges endangered. Within theories with more than one supersymmetry
this scenario could possibly lead to a second important constraint (besides
the discussed sign of $F^2$).

\section{Summary and Conclusions}
In this paper we presented the basic tools needed to test dynamical
symmetry breaking in the context of supersymmetric quantum field theories and
we discussed the application thereof to the simplest interesting model,
$N=1$ SYM.

We showed that $N=1$ SYM comprises an unique source extension being covariant under all
supersymmetries while conserving gauge symmetry. From the fundamental concept of
studying symmetry breaking as hysteresis effect, this source extension alone
can and must be used to answer the question of dynamical supersymmetry
breaking. We explained in detail the connection of our procedure to
thermodynamical limits and classified it in the context of exactly
solvable as well as other field theories: In a four-dimensional theory with a
non-perturbative sector (or any other theory that cannot be solved
axiomatically) this concept is the only one which allows us to
\emph{determine} the ground-state. We discussed in detail why other low energy approximation (some of
them also called effective action or effective Lagrangian but conceptually
different from the quantum effective action) are not suitable tools to answer
the question of dynamical supersymmetry breaking -- nevertheless, after
identifying the correct ground-state from a complete study of thermodynamical
limits (or after extracting it from experimental results) these concepts can be useful to describe the dynamics over this state.

We discussed the relevant thermodynamical limit
explicitly under the assumption that extrinsic supersymmetry is realized on
the effective potential as superspace-geometry on the level of the classical
fields. We discussed in detail the assumption that can and must be made in
order to be able to make any statements but leave the question of dynamical
supersymmetry breaking open. We gave some comments about this construction from the point of
view of perturbation theory. Especially we motivated that unbroken
supersymmetry is rather assumed therein than found as a result.

Comparing semiclassical analysis of QCD
with our result leads to the observation, that the sign of the vacuum energy
(being defined as the vacuum expectation value of the energy-momentum tensor)
of SYM lies in a unphysical region from the point of view of QCD. This raises
the question of possible connections of these two theories. On the
semiclassical level this can be established by means of non-decoupling
theorems and they lead to a surprising conclusion: The specific form of the
superpotential  resulting from supersymmetric non-linear $\sigma$-models implies the existence of a phase transition
separating supersymmetric theories from non-supersymmetric ones. We discussed
this in detail for $N=1$ SYM and found that the phase transition in the gluino
mass must be at $m=0$. However, a supersymmetric theory
protected by a phase transition would not obey minimal thermodynamical
stability conditions and the infrared problem at the supersymmetry
preserving point is not getting removed. Moreover a phase transition leads to
a serious conceptual problem: Lacking understanding of non-perturbative effects,
they are included in all standard approximations by assumption in comparison
with theories realized in nature. Such assumptions based on physical arguments
are clearly unfounded if the theory does not have any connection to a
physically relevant model.

To get an acceptable behavior we conclude that the phase transition does not
exist. The effective potential receives important contributions that cannot be
written in terms of standard
non-linear $\sigma$-models. This is
closely related to the question of the relevance of the auxiliary fields, as the potential of
the latter is now getting changed by non-perturbative effects. This removes
all potential instabilities in the infrared region but supersymmetry is
completely run over by chiral symmetry breaking, confinement and the dynamical formation of a
mass gap. We are not able to show from first principles that supersymmetry
must break dynamically but even assuming the possibility of
conserved supersymmetry in this new scenario has drastic consequences: The low
energy behavior thereof would be completely different than the standard
solutions from effective Lagrangians or instanton calculations: All these
results are based on the assumption that supersymmetry on the level of
classical fields is still realized as an integral over superspace. For our new
solution we have to modify this picture.

At the current status of the discussion we are not able to decide whether $N=1$ SYM
breaks supersymmetry dynamically or not. However the illustration of the
different arguments and restrictions in favor of and against dynamical
supersymmetry breaking unraveled the following point: Within the theoretical
discussion there exists an important difference between theories accessible in
experiments (QCD) and others (SYM or SQCD). For the first class different
approximations to the low energy dynamics using semiclassical and/or momentum
expansion have been very successful. The application of such techniques often
draw upon the known vacuum structure of the theory and thus describe the
dynamics over this structure, only. In supersymmetry this restriction is
obviously impossible and thus it has been tried to use the same techniques for
both, the determination of the vacuum structure and the description of the
dynamics. This ansatz led to a consistent description of many supersymmetric
gauge-theories.

Our discussion shows that these results are nevertheless problematic. All
tools used above factor out important aspects of the low energy dynamics,
namely non-perturbative non-semiclassical effects. Though the consistency of
the standard picture of supersymmetry strongly suggests a coherence of all
these models within the given approximation, we showed that important
non-semiclassical effects can not be excluded, neither by LEEA's, nor by
Instanton calculations or by the Witten index. We thus propose to re-analyze
the structure of the ground-state including the full dynamics of
the system, which is done by using the QEA as fundamental object. Such an
ansatz asks for a completely new interpretation of superspace geometry and of
the role of the auxiliary fields. Many aspects therein are still unclear and
though some promising progress towards an understanding of such models has
been made recently (presented in \cite{bergamin02:1}), a concrete model
describing at least $N=1$ SYM in not yet in sight. Whatever such a model will
look like, we can foresee that it will not be compatible with the LEEA or
effective Lagrangian description of the theory, as the standard superspace
geometry used in the latter approaches can not be relevant in the first
case. In this situation we conclude in analogy to QCD that the QEA must be the
more fundamental object. Therefore in our opinion the LEEA has to conform to
the results from the QEA.

Many other questions are still open. Even on the level of $N=1$ SYM the vacuum
expectation value of the Lagrangian now consists of two different operators
that cannot be separated in a supersymmetry covariant way. However,
independent knowledge about each of them is needed to learn about the
principles of supersymmetry breaking, especially about the goldstino coupling. In
more complicated models this becomes even more important. As an example
confinement could be realized in $N=2$ SYM by trilinear Yukawa-like
condensates, but supersymmetry connects the corresponding operators to the
scalar potential and to $F_{\mu \nu} F^{\mu \nu}$. When coupling matter fields
to $N=1$ or when breaking $N=2$ explicitly to $N=1$ the role of the matter
auxiliary field $F$ must be studied. The latter can no longer be eliminated
naively: In $N=2$ $F$ and $D$ are related by the internal $\su(2)$ symmetry
and more generally the elimination leads to unacceptable decoupling
behaviors. An interpretation could possibly be found by studying the structure of the
energy-momentum tensor: Therein the $F$ fields are important for the
correct breaking of superconformal invariance and could play a similar role as
the $D$ field in $N = 1$ SYM.


\section*{Acknowledgement}
It is a pleasure to thank all members of the theory group in Bern for
enlightening discussions. Especially I would like to thank P. Minkowski for
his support and for numerous interesting discussions and Ch.~Rupp for useful
comments on perturbative aspects of supersymmetry. I am also grateful to the members of the SLAC theory
group and of the Institute for Theoretical Physics in Leipzig for their warm
hospitality. Especially I would like to acknowledge illuminating discussions
with K. Sibold and with M. Peskin. Finally I am grateful to E. Kraus for
useful comments about perturbative SYM with local coupling constants.


\appendix
\bibliography{biblio}
\end{document}